\documentclass[onecolumn,apjl,numberedappendix]{openjournal}
\usepackage{xcolor}
\usepackage{textgreek}
\usepackage[utf8]{inputenc}
\usepackage[english]{babel}
\usepackage{amsmath}
\usepackage{color,colortbl}
\usepackage{tensind}
\tensordelimiter{?}
\DeclareGraphicsExtensions{.bmp,.png,.jpg,.pdf}
\usepackage{verbatim}
\usepackage[normalem]{ulem}
\usepackage{orcidlink}
\usepackage{soul}
\usepackage{booktabs}
\usepackage{array}
\usepackage{enumitem}
\usepackage{graphicx}
\usepackage{fontawesome5}

\definecolor{linkcolor}{HTML}{F21A00}

\usepackage{hyperref}
\hypersetup{
    unicode=true,
    colorlinks=true,
    linkcolor=linkcolor,
    citecolor=linkcolor,
    urlcolor=linkcolor,
    filecolor=linkcolor,
}

\newcommand{\pkgfont}{\texttt}
\newcommand{\pkg}[2][]{%
  \if\relax\detokenize{#1}\relax
    \pkgfont{#2}%
  \else
    \href{#1}{\pkgfont{#2}}%
  \fi
}

\newcommand{\derivkit}{\texttt{DerivKit}}

\newcommand{\emcee}{\texttt{emcee}}

\newcommand{\dali}{\text{\scshape dali}}
\newcommand{\map}{\text{\scshape map}}

\newcommand{\mcmc}{\text{\scshape mcmc}}
\newcommand{\mcse}{\text{\scshape mcse}}

\newcommand{\num}{\text{\scshape num}}
\newcommand{\fish}{\text{\scshape f}}

\newcommand{\meantheta}{\ensuremath{\bar{\theta}}}
\newcommand{\maptheta}{\ensuremath{\hat{\theta}}}

\newcommand{\meanshift}{\ensuremath{\Delta\bar{\theta}}}
\newcommand{\mapshift}{\ensuremath{\Delta\hat{\theta}}}

\newcommand{\gaussshift}{\ensuremath{\Delta\theta_{\mathrm{G}}}}
\newcommand{\shiftgaussian}{\ensuremath{\delta_{\mathrm{G}}}}
\newcommand{\shiftnongaussian}{\ensuremath{\delta_{\mathrm{NG}}}}

\newcommand{\linbias}{\ensuremath{\Delta\theta_{\mathrm{F}}}}

\newcommand{\ddata}{\ensuremath{d_{\mathrm{data}}}}

\newcommand{\dd}{\mathop{}\!\mathrm{d}}

\newcommand{\sfrac}[2]{{\scriptstyle\frac{#1}{#2}}}

\newcommand{\avg}[1]{\langle #1 \rangle}

\newcommand{\transp}{^{\mathrm{T}}}

\newcommand{\llikelihood}{\ensuremath{\mathcal{L}}}

\newcommand{\eg}{\emph{e.g.}}
\newcommand{\ie}{\emph{i.e.}}

\newcommand{\nasa}{\textsc{nasa}}
\newcommand{\doe}{\textsc{doe}}

\newcommand{\lofar}{\textsc{lofar}}
\newcommand{\uk}{\textsc{uk}}
\newcommand{\uksrc}{\textsc{uksrc}}

\newcommand{\stagefour}{Stage~\textsc{iv}}

\makeatletter
\def\@eapj@cap@font{\normalfont}
\makeatother

\graphicspath{ {./figures/} }

\begin{document}
\title{Non-Gaussian Parameter Bias: Formalism and Validation}

\shorttitle{Non-Gaussian Parameter Bias}
\shortauthors{\v{S}ar\v{c}evi\'{c} \& Van der Wild \textit{et al.}}

% Author list is here
\author{Nikolina \v{S}ar\v{c}evi\'{c}\orcidlink{0000-0001-7301-6415}}
\email{nikolina.sarcevic@gmail.com}
\affiliation{\textit{Department of Physics, Duke University, Science Dr, Durham, NC 27710, USA}}

\author{Matthijs van der Wild\orcidlink{0000-0002-3949-3063}}
\email{matthijs@vanderwild.com}
\affiliation{\textit{Department of Physics, Durham University, Lower Mountjoy, South Rd, Durham DH1 3LE, UK}}
\affiliation{\textit{Department of Physics, Duke University, Science Dr, Durham, NC 27710, USA}}

\author{Elena Sellentin\orcidlink{0009-0002-2655-3458}}
\affiliation{\textit{Leiden Observatory, Leiden University, PO Box 9513, NL-2300 RA Leiden, the Netherlands}}

% Abstract
\begin{abstract}
We present a framework for propagating systematic data mismatches into parameter biases and posterior deformations beyond the Gaussian Fisher approximation.
Using the Derivative Approximation for Likelihoods (\dali), we derive analytic expressions for the maximum a posteriori (\map) point and posterior mean, including a semi-analytic response expansion in systematic amplitude.
We validate the framework against direct numerical calculations and Markov Chain Monte Carlo (\mcmc) sampling in a controlled nonlinear two-parameter benchmark up to a covariance-weighted data-space mismatch of $\ddata=5$.
At $\ddata=5$, Fisher predicts the wrong direction of the shift in one parameter and differs from the numerical result by a Fisher-metric distance of $23.1$, whereas the analytic \dali\ \map\ prediction remains within $0.091$, with componentwise errors of $0.04\%$ and $0.46\%$.
The response expansion reproduces the posterior mean with errors of $0.02\%$ and $0.18\%$, while the \dali\ posterior closely reproduces the displacement and deformation of the fully sampled nonlinear posterior.
Mismatches of equal significance but different directions produce markedly different parameter shifts and posterior deformations, while multiple additive mismatches can produce nonadditive parameter responses through nonlinear inference.
By reconstructing the posterior under systematic mismatch, the framework captures the nonlinear response and non-Gaussian geometry of parameter bias: how mismatches move and reshape the posterior, why their direction matters, and how their combined effects can be dissected to understand the interplay between multiple systematics.
\end{abstract}

% Write your keywords here

\maketitle

{\small \tableofcontents}
\vspace{1cm}

% Main paper sections
\section{Introduction}
\label{sec:intro}

A central challenge in cosmological analyses is to propagate observational and modelling systematics into parameter uncertainties and associated biases, particularly when the posterior departs from Gaussianity.
In the Gaussian approximation, parameter biases induced by data--model mismatches are commonly estimated using the Fisher bias formalism \citep{HUTERER2005369, Taylor_2007, Amara_2008}.
This approach relies on a locally quadratic log-likelihood and a linearized response of the observables to model parameters.
Previous work has noted that, for sufficiently large biases, changes in the likelihood curvature can invalidate this approximation and motivate either higher-order expansions or evaluation of the full likelihood \citep{Kitching_2009}.
These assumptions can break down in the presence of nonlinear parameter dependencies or non-Gaussian likelihood structure, for which the Fisher approximation can fail to provide an accurate description of parameter constraints and posterior structure \citep{Wolz_2012, Park_2023, Ryan_2023}.

The Derivative Approximation for Likelihoods (\dali) \citep{Sellentin_2014} provides a framework to move beyond the Gaussian limit by capturing nonlinear parameter dependence through a higher-order expansion of the log-likelihood.
The \dali\ framework has subsequently been applied to non-Gaussian cosmological forecasting, including weak-lensing applications \citep{Sellentin_2015, Samario_Nava_2025}.
More general expressions for relating best-fit parameter shifts induced by inaccurate modelling of the observable have been developed within the Fisher framework \citep{Bernal_2020}.
However, an analogous treatment of systematic data--model mismatches beyond the Gaussian approximation, in which the mismatch is propagated through the higher-order posterior itself and the corresponding values of the inferred parameters, has not yet been developed.
Here, we present this extension within the \dali\ framework.

Current \stagefour\ cosmological surveys, including the Legacy Survey of Space and Time at the Vera C. Rubin Observatory \citep{lsst_stageiv_ivezic}, the \textit{Euclid} mission \citep{euclid_stageiv_scaramella}, \nasa\ Roman Space Telescope \citep{roman_stageiv_spergel}, the Dark Energy Spectroscopic Instrument \citep{desi_stageiv_abareshi}, and the Simons Observatory \citep{simonsobs_stageiv}, increasingly rely on small-scale information and incorporate a large number of nonlinear astrophysical and observational systematics.
Models of intrinsic alignments, galaxy bias, baryonic feedback, and photometric redshift uncertainties introduce substantial nuisance complexity into cosmological analyses, with several of these effects requiring nonlinear modelling as increasingly small scales are included (\eg\ \citealt{Chisari_ia_review, Robertson_2026, truttero2026balancingbiasbaryonsscale}).
In this regime, the classical Gaussian treatments of parameter biases can become unreliable, while fully sampling the posterior across all combinations of systematics and modelling choices can be computationally prohibitive \citep{Ryan_2023}.
Even when feasible, for example via Markov Chain Monte Carlo (\mcmc) methods, posterior sampling does not, by itself, isolate how individual effects contribute to the inferred shifts and deformations.
A derivative-based treatment beyond the Gaussian approximation provides a complementary route, capturing leading non-Gaussian corrections and the interaction of multiple systematics without requiring full posterior sampling for each configuration.

Beyond providing a corrected bias estimate, the nonlinear \dali\ bias also provides a way to characterise the geometry of the likelihood beyond the local Fisher approximation.
By relating higher-order derivative tensors to the resulting nonlinear parameter response, prescribed systematic mismatches can be followed as they displace the inference away from the local Gaussian regime.
For multiple systematics, departures from additivity further isolate contributions to the inferred shift that arise from their joint propagation through the nonlinear inference map.
The central aim of this work is to develop parameter bias forecasting beyond the Gaussian approximation while retaining the efficiency and interpretability of a derivative-based approach.
A key consequence of moving beyond Gaussianity is that the maximum a posteriori (\map) point and posterior mean no longer generally coincide.
The notion of a single parameter bias therefore separates naturally into distinct \map\ and posterior mean shifts, which probe different properties of the posterior and require separate predictions.

In this work, we develop a systematic approximation scheme and use it to derive both quantities within the \dali\ framework.
For the \map, the higher-order likelihood expansion yields a systematic \dali\ generalisation of the Fisher bias equation, whose leading-order limit recovers the standard Fisher result.
For the posterior mean, we derive a local analytic approximation and develop a complementary expansion in the amplitude of the systematic mismatch.
The latter retains the complete non-Gaussian \dali\ posterior as its reference measure.
These objects provide a unified framework for predicting how systematic data mismatches propagate into both local shifts of the posterior maximum and global shifts of its mean.
The resulting formalism is implemented using the \derivkit\ \citep{derivkit} ecosystem~\href{https://github.com/derivkit/derivkit}{\faGithub}\footnote{\url{https://github.com/derivkit/derivkit}}, which provides the derivative-based computational tools for evaluating the required Fisher and higher-order \dali\ tensors and their associated bias predictions.

We validate the predictions using a controlled nonlinear benchmark, varying the systematic mismatch to probe regimes in which non-Gaussian posterior structure becomes increasingly important.
The benchmark is simple enough to permit direct numerical determination of the \map\ and posterior moments, with independent comparisons to \mcmc\ sampling, while retaining nonlinear features representative of cosmological inference problems.
This provides a direct test of the analytic and response predictions across a range of systematic amplitudes and allows us to assess where they remain accurate relative to the standard Fisher approximation.

Specifically, we
\begin{enumerate}
    \item[(i)] derive the non-Gaussian \dali\ bias equation and show explicitly how the standard Fisher bias is recovered at leading order;
    \item[(ii)] establish the distinction between systematic shifts of the posterior \map\ and posterior mean beyond the Gaussian approximation;
    \item[(iii)] derive analytic \dali\ predictions for the \map\ and posterior mean shifts induced by systematic data--model mismatches in the presence of nonlinear parameter dependence;
    \item[(iv)] develop a systematic semi-analytic response expansion for the posterior mean that retains the complete non-Gaussian \dali\ posterior as its reference measure; and
    \item[(v)] validate these predictions against direct numerical calculations and \mcmc\ sampling across a range of systematic amplitudes in the nonlinear benchmark model.
\end{enumerate}

The remainder of this manuscript is organised as follows.
Section~\ref{sec:linear_to_nonlinear} presents the \dali\ bias formalism and its relation to the standard Fisher bias.
Section~\ref{sec:map_and_mean_shifts} distinguishes the \map\ and posterior mean shifts and derives their corresponding non-Gaussian predictions.
Section~\ref{sec:modelling_and_inference} introduces the nonlinear benchmark and numerical inference strategy used to validate the formalism.
Section~\ref{sec:posterior_reconstruction} presents the systematic parameter shifts and their relation to the nonlinear posterior response as the mismatch is increased.
We conclude in Section~\ref{sec:discussion_and_summary} with a discussion, summary, and outlook.
Detailed derivations, numerical convergence tests, and additional validation results are collected in the appendices.

\section{From Fisher bias to non-Gaussian posterior response}
\label{sec:linear_to_nonlinear}
Parameter biases in cosmological analyses arise when the assumed model does not accurately describe the data.
When the model fails to incorporate systematic effects, the predicted posterior for cosmological parameters is shifted with respect to the true posterior%
\footnote{
    This work appeals to the principle of indifference and  adopts a uniform prior for simplicity.
    The numerical insensitivity of the reported posterior results to the adopted finite parameter bounds is verified in Appendix~\ref{app:numerical-validation}.
}.
For a Gaussian data likelihood, the data vector $d$ is characterised by a covariance matrix%
\footnote{
In this work, $C$, and hence $M$, is assumed to be independent of the model parameters.
}
$C$ with precision matrix $M \equiv C^{-1}$.
Given a model prediction function $m$, the effect of a data--model mismatch is captured by the Fisher bias formalism \citep{HUTERER2005369, Taylor_2007, Amara_2008, sersante2026invertingfisherbiasesfast}, which relates the parameter shift to the projection of the mismatch onto the linear response of the observables.
The latter is captured by the data--model mismatch vector
\begin{equation}
    \label{eq:mismatch_vector}
    \Delta m \equiv \langle d \rangle - m(\theta_0),
\end{equation}
where the average denotes an ensemble expectation over realisations of the statistical noise.
It therefore isolates the systematic displacement of the mean data vector from the model prediction, rather than the statistical fluctuation of a particular realisation.
Here $d$ is the data vector, $m(\theta)$ is the model prediction evaluated at parameters $\theta$, and $\theta_0$ represents the fiducial parameter values.
The (linear) Fisher bias is given by
\begin{equation}
    \label{eq:fisher_bias}
    \linbias^a = F^{ab} b_b,
\end{equation}
with $F^{ab}$ the inverse Fisher matrix (\ie\ the parameter covariance matrix in the Gaussian approximation), and%
\footnote{
An upright uppercase superscript ${\rm T}$ denotes the transpose operator.
}
\begin{equation}
    \label{eq:linear_response}
    b_a = \Delta m\transp M m_a
\end{equation}
encodes the linear response of the model to the data--model mismatch.
The function $m_a = \partial m / \partial \theta^a$ is the derivative of the model with respect to parameter $\theta^a$ evaluated at $\theta_0$.
Repeated parameter indices are summed over.
Higher-order tensors can be defined similarly, with more indices denoting higher-order derivatives.

While widely used, this expression assumes that the model response from a data--model mismatch is linear.
It also assumes that higher-order derivatives of the model functions can be neglected.
However, as discussed in the Introduction, realistic cosmological models have complex nuisance effects which nonlinearly influence the underlying model, including baryonic feedback, intrinsic alignments, and galaxy bias.
Such systems can exhibit significant higher-order parameter dependence, making higher-order derivatives of the model important.
These derivatives introduce nonlinear corrections to the bias and, subsequently, allow different systematic effects to interact in parameter space.
The parameter shift induced by multiple systematics is therefore not guaranteed to equal the sum of the shifts induced by each systematic individually.
The linear Fisher formalism does not capture these interactions.
It can therefore misestimate both the magnitude and the direction of the parameter shift.

To go beyond this linear approximation, one must include higher-order derivatives of the model function to account for its nonlinear dependence on the model parameters.
Retaining the second derivative of the model while keeping the posterior approximation Gaussian yields a second correction $b_{ab} = \Delta m\transp M m_{ab}$, and the shift due to the bias reads
\begin{align}
    \gaussshift &=
    (1 - F^{-1}B)^{-1}\linbias\nonumber
    \\
    &= \linbias + F^{-1}B\linbias + O(\vert\Delta m\vert^3).
    \label{eq:gaussian_shift}
\end{align}
Here $B$ is the matrix with components $B_{ab} = b_{ab}$ and $F^{-1}$ is the matrix with components $F^{ab}$.
This correction is derived in Appendix~\ref{ap:ap1}.

Capturing non-Gaussian posterior structure arising from nonlinear model dependence requires terms beyond the second-order Taylor expansion.
One way of doing this consistently is through the derivative approximation for likelihoods (\dali) method \citep{Sellentin_2014}, which reweights terms in the series expansion of the log-likelihood based on the order of derivatives contained in each term.
When considering derivatives of at most order 3, the \dali\ approximation includes the Fisher matrix%
\footnote{While not conventionally considered to be part of the \dali\ scheme, the Fisher matrix occupies the \dali\ singlet.},
two doublet tensors $D^1$ and $D^2$, and three triplet tensors $T^1$, $T^2$, and $T^3$.
They take the form
\begin{align}
    F_{ab} &= m_a\transp M m_b,\nonumber\\
    D^1_{abc} &= m_{(a}\transp M m_{bc)},\nonumber\\
    D^2_{abcd} &= m_{(ab}\transp M m_{cd)},\nonumber\\
    \label{eq:dali-definition}
    T^1_{abcd} &= m_{(a}\transp M m_{bcd)},\\
    T^2_{abcde} &= m_{(ab}\transp M m_{cde)},\nonumber\\
    T^3_{abcdef} &= m_{(abc}\transp M m_{def)},\nonumber
\end{align}
and quantify the model response to parameter displacements about the fiducial point.
The parentheses in the subscripts denote symmetrisation over the enclosed indices:
\begin{equation}
T_{(a_1\ldots a_n)} = \sfrac{1}{n!} (\text{sum of all permutations of the indices of }T).
\end{equation}
In the \dali\ approximation, the log-likelihood $\llikelihood$ can be approximated in a region about $\theta_0$ as
\begin{align}
\llikelihood(d\vert\theta_0 + \delta) =&
        \llikelihood(d\vert\theta_0)
            - \sfrac{1}{2} F_{ab} \delta^a \delta^b
            - \sfrac{1}{2} D^1_{abc} \delta^a \delta^b \delta^c
            - (
                \sfrac{1}{3!} T^1_{abcd}
                + \sfrac{1}{8} D^2_{abcd}
            ) \delta^a \delta^b \delta^c \delta^d
            \nonumber\\&\quad
            - \sfrac{1}{12} T^2_{abcde} \delta^a \delta^b \delta^c \delta^d \delta^e
            - \sfrac{1}{72} T^3_{abcdef} \delta^a \delta^b \delta^c \delta^d \delta^e \delta^f
            \nonumber\\&\quad
            + b_a \delta^a
            + \sfrac{1}{2} b_{ab} \delta^a \delta^b
            + \sfrac{1}{3!} b_{abc} \delta^a \delta^b \delta^c.
    \label{eq:likelihood}
\end{align}
The additional terms, which encode the bias, are defined in analogy to $b_a$:
\begin{equation}
    \label{eq:bias-definition}
    b_{a_1\ldots a_n} = \Delta m\transp M m_{a_1\ldots a_n}.
\end{equation}
These tensors modify the \dali\ contributions to the log-likelihood due to data--model mismatch.

So far, $\gaussshift$ denoted a generic local parameter displacement obtained from \eqref{eq:gaussian_shift}.
To characterise the posterior response to a data--model mismatch, we now distinguish two particular estimators.
These are the maximum a posteriori (\map) estimator $\maptheta$, defined as the point that maximises the posterior density, and the posterior mean $\meantheta \equiv \mathbb{E}[\theta]$.
For Gaussian posteriors these are identical.
Beyond the Gaussian approximation, however, different posterior estimators need not coincide.
Section~\ref{sec:map_and_mean_shifts} derives the two shifts explicitly.

\section{\map\ and posterior mean shifts}
\label{sec:map_and_mean_shifts}

\subsection{Maximum a posteriori}

The \map\ shift $\mapshift$ is defined by the condition that the gradient of the log-posterior vanishes:
\begin{equation}
\label{eq:stationarity-condition}
\partial_a \llikelihood(\theta_0 + \mapshift) = 0.
\end{equation}
While \eqref{eq:stationarity-condition} does not generally admit a closed-form solution for non-Gaussian posteriors, it can be solved perturbatively to yield the \map\ bias shift as an expansion around the Gaussian solution.
This leads to a solution of the form
\begin{equation}
\label{eq:dalibias_general}
    \mapshift^a = \gaussshift^a + \sum_{n=2}^{\infty} \shiftnongaussian{}_{n}^a,
\end{equation}
where $\gaussshift$ is given by \eqref{eq:gaussian_shift} and can similarly be decomposed in powers of $\Delta m$.
Here the subscript $n$ denotes order in the data--model mismatch, such that
\begin{equation}
    \shiftnongaussian{}_{n} \propto \vert\Delta m\vert^n.
\end{equation}
This ordering refers to powers of the data--model mismatch and should not be confused with the order of the local \dali\ expansion.
The derivation of this expansion is shown in Appendix~\ref{app-sec:map}.
The expressions for $\shiftnongaussian{}_{n}$ up to order $n=3$ are
\begin{align}
    \shiftnongaussian{}_{2}^a &=
        - \sfrac{3}{2} F^{ab} D^1_{bcd} \shiftgaussian{}_1^c \shiftgaussian{}_1^d,
\\
    \shiftnongaussian{}_{3}^a &=
        F^{ab} \Big[
            b_{bc} \shiftnongaussian{}_{2}^c
            + \sfrac{1}{2} b_{bcd} \shiftgaussian{}_{1}^c\shiftgaussian{}_{1}^d
            - 3 D^1_{bcd}\shiftgaussian{}_1^c\left(
                \shiftgaussian{}_2^d
                + \shiftnongaussian{}_2^d
            \right)
\\&\qquad
            - \sfrac{1}{6}(
                4 T^1_{bcde}
                + 3 D^2_{bcde}
            ) \shiftgaussian{}_1^c\shiftgaussian{}_1^d\shiftgaussian{}_1^e
        \Big],
\end{align}
The full derivation in Appendix~\ref{app-sec:map} includes contributions up to $O(\vert\Delta m\vert^5)$, which is sufficient to incorporate the complete effects of the \dali\ triplet.

\subsection{Analytic expansion of the posterior mean}

In contrast, the posterior mean shift $\meanshift$ measures the displacement of the posterior mean from the reference parameter point $\theta_0$,
\begin{equation}
\label{eq:mean_bias_definition_main}
    \meanshift
    \equiv
    \meantheta - \theta_0,
\end{equation}
where $\theta_0$ is the point about which the \dali\ expansion is constructed, and the overbar denotes an expectation value with respect to the full posterior,
\begin{equation}
\label{eq:expectation-value}
\bar{X} = \int X\, p(\theta\vert d)\,\dd^{N_{\textrm par}}\theta.
\end{equation}
$N_{\textrm par}$ denotes the number of parameters.
As stated in Section \ref{sec:linear_to_nonlinear}, for a Gaussian posterior the shift in the mean is equal to the shift in the \map\ and is given by \eqref{eq:gaussian_shift}.
Beyond the Gaussian approximation, $\meanshift$ need not coincide with $\mapshift$.
It is shown in Appendix~\ref{app-sec:mean} that the difference between the two is
\begin{equation}
\label{eq:mean_bias_main_scheme}
    \meanshift^a
    =
    \mapshift^a
    - \sfrac{1}{2} C^3_{ijk} (C^2)^{ia}(C^2)^{jk}
    - \sfrac{1}{8} C^5_{ijklm}(C^2)^{ia}(C^2)^{jk}(C^2)^{lm},
\end{equation}
up to the order of the \dali\ triplet.
The $C^n$ coefficients are given by
\begin{equation}
\begin{aligned}
    C^n_{i_1\ldots i_n}
    &= \sum_{r=n}^6 \sfrac{1}{(r-n)!}
    c^r_{i_1\ldots i_n\ldots i_r}
    \mapshift^{i_{n+1}}\ldots\mapshift^{i_r},
    \\
    c^2_{ab} &= F_{ab} - b_{ab}, \\
    c^3_{abc} &= 3 D^1_{abc} - b_{abc}, \\
    c^4_{abcd} &= 4 T^1_{abcd} + 3 D^2_{abcd}, \\
    c^5_{abcde} &= 10 T^2_{abcde}, \\
    c^6_{abcdef} &= 10 T^3_{abcdef}.
\end{aligned}
\end{equation}
This, too, is derived in Appendix~\ref{ap:ap1}.

Equation~\eqref{eq:mean_bias_main_scheme} gives the displacement of the posterior mean relative to the expansion point $\theta_0$.
For a non-Gaussian posterior this displacement can be nonzero even in the absence of a data--model mismatch.
The contribution induced specifically by the data--model mismatch is therefore obtained by comparing the posterior mean in the presence of the mismatch with that of the corresponding unbiased posterior.

\subsection{Semi-analytic expansion of the posterior mean}
\label{subsec:response_mean_shift}

The preceding perturbative approximation is only correct up to the level of the \dali\ triplet and, therefore, does not retain the complete exponential structure of the unbiased \dali\ posterior.
To account for this, we present a semi-analytic expansion about the unbiased \dali\ posterior, retaining its intrinsic non-Gaussian structure in the reference distribution and expanding only its response to the data--model mismatch.
The two approaches are complementary: the analytic expression provides a closed local relation between the \map\ and mean directly in terms of the \dali\ tensors, while the semi-analytic response formulation moves beyond this purely local form by retaining the complete non-Gaussian unbiased \dali\ posterior as its reference distribution.

As shown in Appendix~\ref{app-sec:mean-response}, to cubic order in the data--model mismatch parameter $\Delta m$, the mean shifts as
\begin{align}
\label{eq:semi-analytic}
\meantheta^a &= \meantheta^a_0
            + \avg{(\theta^a - \meantheta^a_0)(B-\avg{B})}
            \nonumber\\&\qquad
            + \sfrac{1}{2} \avg{(\theta^a - \meantheta^a_0)(B - \avg{B})^2}
            \nonumber\\&\qquad
            + \sfrac{1}{3!} \avg{(\theta^a - \meantheta^a_0)(B - \avg{B})^3}
            \nonumber\\&\qquad
            - \sfrac{1}{2} \avg{(\theta^a - \meantheta^a_0)(B - \avg{B})}\avg{(B -\avg{B})^2}.
\end{align}
Here, the scalar function $B$ is defined as
\begin{equation}
    \label{eq:app-bias-factor}
    B(\theta)
    =
    \sum_n \sfrac{1}{n!} b_{a_1 \ldots a_n} \theta^{a_1} \dots \theta^{a_n}
\end{equation}
and should not be confused with the shorthand used in \eqref{eq:gaussian_shift}.
The $b$ tensors are defined as in \eqref{eq:bias-definition}.
The expectation values are computed relative to the posterior obtained from \eqref{eq:likelihood} with $\Delta m = 0$.
Its mean has been denoted as $\meantheta_0$.
From \eqref{eq:semi-analytic} one can straightforwardly read off the systematic posterior mean shift relative to the unbiased posterior.
Unlike the preceding analytic mean approximation \eqref{eq:mean_bias_main_scheme}, the reference measure retains the complete non-Gaussian \dali\ posterior, while the expansion is performed only in the mismatch vector $\Delta m$.
In practice, we retain terms through cubic order in the data--mismatch vector, which is sufficient for the systematic amplitudes considered here, as verified against direct numerical evaluation of the \dali\ posterior in Section \ref{sec:posterior_reconstruction}.

\section{Modelling and inference}
\label{sec:modelling_and_inference}

We implemented the \dali\ parameter bias formalism using the derivative and inference tools provided by \derivkit\ and applied it to a controlled nonlinear benchmark.
The model was deliberately constructed to exhibit a strongly nonlinear parameter-to-data mapping and a highly non-Gaussian posterior while remaining numerically tractable.
Because the benchmark model is cubic, the triplet \dali\ representation reproduces its full nonlinear parameter dependence up to numerical derivative errors.
The benchmark therefore provides a controlled setting in which to isolate and validate the bias-response predictions themselves, without conflating their accuracy with truncation error in the \dali\ representation of the likelihood.
Two distinct systematic data mismatches perturb different combinations of the observables.
We investigated each independently and then simultaneously, allowing us to compare the posterior response to the individual and combined systematics.

\subsection{Nonlinear benchmark}
\label{subsec:nonlinear_benchmark}

The Fisher and \dali\ bias predictions derived in Section~\ref{sec:map_and_mean_shifts} are defined in terms of model derivatives evaluated at the fiducial parameter point $\theta_0$.
The triplet \dali\ posterior uses model derivatives through third order, while the perturbative \map\ prediction includes terms through fifth order in the data–model mismatch, as required to incorporate the complete triplet-level \dali\ contributions.

For the numerical benchmark, we parameterised the mean data vector as $\bar d(A)=m_0+A\,s$, where $m_0=m(\theta_0)$, $s$ specified a fixed systematic template in data space, and $A$ controlled its amplitude.
Given an estimator $E$, let $\theta_E(A)$ denote its value for the posterior obtained at amplitude $A$, and let $\theta_E(0)$ denote its value for the uncontaminated posterior.
We then define its systematic shift relative to the uncontaminated posterior as
\begin{equation}
    \Delta\theta_E(A)
    =
    \theta_E(A)-\theta_E(0).
    \label{eq:systematic-shift}
\end{equation}
This definition is particularly important for the posterior mean, since a non-Gaussian uncontaminated posterior can have $\bar{\theta}(0)\neq\theta_0$.
The systematic shift therefore isolates the response to the injected mismatch from any intrinsic offset of the uncontaminated posterior mean from the fiducial point.

The benchmark consists of two parameters $(u,v)$ and produces a strongly non-Gaussian posterior with a pronounced curved parameter degeneracy.
Its low-dimensional parameter space allows the posterior moments to be evaluated accurately by direct numerical integration while retaining the nonlinear structure required to stress the Fisher and higher-order \dali\ predictions.
The main benchmark settings are summarised in Table~\ref{tab:benchmark-settings}.

\begin{table}[h!]
    \centering
    \caption{Configuration of the nonlinear benchmark model.}
    \label{tab:benchmark-settings}
    \begin{tabular}{ll}
        \hline
        Quantity & Definition or value \\
        \hline
        Parameters
        & $\theta=(u,v)$ \\
        Fiducial point
        & $\theta_0=(u_0,v_0)=(0.15,0.05)$ \\
        Parameter offsets
        & $\Delta u=u-u_0,\quad \Delta v=v-v_0$ \\
        First observable
        & $m_1=1+1.4\Delta u+0.15\Delta v+0.20\Delta u^2+0.06\Delta u\Delta v$ \\
        Second observable
        & $m_2=0.8+10(\Delta v+1.8\Delta u^2-0.3\Delta u^3+0.20\Delta u\Delta v)$ \\
        Fiducial prediction
        & $m_0=(1,0.8)$ \\
        Fractional uncertainty
        & $f=0.10$ \\
        Correlation coefficient
        & $\rho=0.3$ \\
        Covariance
        & $C=\begin{pmatrix}0.0100&0.0024\\0.0024&0.0064\end{pmatrix}$ \\
        Prior
        & $\theta\in[-2,2]^2$ \\
        $\alpha$ template
        & $s_\alpha=(0.020,-0.003)$ \\
        $\beta$ template
        & $s_\beta=(0.003,0.016)$ \\
        Systematic configurations
        & $\alpha,\ \beta,\ \alpha+\beta$ \\
        Amplitude range
        & $A=0,1,\ldots,20$ \\
        \hline
    \end{tabular}
\end{table}

Because the model is polynomial through cubic order, its third-order Taylor expansion about $\theta_0$ is exact.
With the fixed covariance used here, the resulting triplet \dali\ likelihood therefore reproduces the full nonlinear likelihood, up to numerical derivative errors.
With the same prior, this equivalence also holds for the posterior.
The benchmark consequently isolates the accuracy of the perturbative \map\ and posterior-mean shift predictions when the underlying \dali\ likelihood itself is exact.
The first observable provides a comparatively broad constraint on the parameters.
The second constrains a nonlinear combination of the two parameters, producing the curved degeneracy and associated non-Gaussian posterior structure.

The observational covariance is constructed using a fractional uncertainty $f=0.10$ for each observable, with $\sigma_i=f|m_{0,i}|$, and a correlation coefficient $\rho=0.3$.
The covariance is held fixed throughout the analysis.
The prior is taken to be uniform over the bounds listed in Table~\ref{tab:benchmark-settings}.
The sensitivity of the benchmark results to these bounds is tested in Appendix~\ref{subsec:posterior_validity_and_bounds}, where comparison with the larger domain $[-12,12]^2$ shows no resolved change in the posterior mass or moments at the numerical precision of the calculation.

We define two additive systematic templates $s_\alpha$ and $s_\beta$ that specify fixed directions and scales in data space, while $A$ controls the amplitude applied to each template.
The templates are specified directly in data units rather than normalised in the covariance metric; consequently, $A$ is specific to the chosen template normalisation, while the corresponding covariance-weighted mismatch magnitude is quantified below by $\ddata$.
The $\alpha$ systematic primarily perturbs the first observable, whereas the $\beta$ systematic primarily perturbs the second observable that constrains the curved parameter combination.
Their distinct directions in data space probe different regions of the nonlinear parameter-to-data mapping and consequently generate different parameter responses.

For a single systematic $X\in\{\alpha,\beta\}$, we use
\begin{equation}
    \bar d_X(A)
    =
    m_0
    +
    A\,s_X.
\end{equation}
We first activate each systematic independently, varying its amplitude while setting the other systematic to zero.
For the combined configuration, both systematics are activated with a common amplitude,
\begin{equation}
    \bar d_{\alpha+\beta}(A)
    =
    m_0
    +
    A\left(s_\alpha+s_\beta\right).
\end{equation}
The two systematic templates are exactly additive in data space.
Any nonadditive response of the inferred parameters in the combined case therefore arises from the nonlinear posterior response rather than from an explicit interaction between the injected systematics.
We consider the $\alpha$, $\beta$, and $\alpha+\beta$ configurations over the amplitude range $A=0,1,\ldots,20$.
The zero-amplitude case corresponds to the uncontaminated data vector.
Varying $A$ therefore scans the posterior response along each fixed systematic template, from the uncontaminated case into a regime of substantial data--model mismatch; the physical size of that mismatch is characterised by $\ddata$ rather than by the numerical value of $A$ itself.

For each systematic configuration, we define the injected mismatch as
\begin{equation}
    \Delta m_X(A)
    =
    \bar d_X(A)-m_0,
\end{equation}
where $X\in\{\alpha,\beta,\alpha+\beta\}$.
We quantify the magnitude of this mismatch using its covariance-weighted norm, equivalent to the Mahalanobis distance \citep{Mahalanobis1936},
\begin{equation}
    \ddata
    =
    \sqrt{
        \Delta m_X(A)\transp
        C^{-1}
        \Delta m_X(A)
    }.
\end{equation}
For a fixed template $s_X$, this becomes
\begin{equation}
    \ddata
    =
    A
    \sqrt{
        s_X\transp
        C^{-1}
        s_X
    },
\end{equation}
so the numerical value of $A$ depends on the normalisation chosen for $s_X$.
Equivalently, defining the covariance-normalised direction
\begin{equation}
    \hat{s}_X
    =
    \frac{
        s_X
    }{
        \sqrt{
            s_X\transp
            C^{-1}
            s_X
        }
    },
    \qquad
    \hat{s}_X\transp
    C^{-1}
    \hat{s}_X
    =
    1,
\end{equation}
the mismatch can be written as
\begin{equation}
    \Delta m_X(A)
    =
    \ddata\,\hat{s}_X.
\end{equation}
Therefore, $\ddata$ provides a normalisation-independent, dimensionless measure of the mismatch magnitude relative to the observational covariance, while $\hat{s}_X$ specifies its direction in data space.

\subsection{Inference strategy}
\label{subsec:inference_strategy}

For every systematic configuration and amplitude, we evaluated both the \dali\ posterior and the posterior of the original nonlinear model.
The numerical \dali\ \map\ was obtained by direct optimisation of the \dali\ posterior, while its posterior mean was computed by direct numerical integration.
For the flat prior adopted here, the posterior is proportional to the likelihood over the integration domain, so that
\begin{equation}
    \label{eq:numerical-posterior-mean}
    \meantheta^a
    =
    \frac{
        \int \theta^a \exp[\llikelihood(d\vert\theta)]\,\dd^{N_{\textrm par}}\theta
    }{
        \int \exp[\llikelihood(d\vert\theta)]\,\dd^{N_{\textrm par}}\theta
    }.
\end{equation}
These calculations provided the reference for validating the corresponding analytic \dali\ \map\ and posterior-mean predictions \eqref{eq:dalibias_general} and \eqref{eq:mean_bias_main_scheme}.
For the posterior mean, we additionally compared with the systematic response expansion \eqref{eq:semi-analytic}.
Independently, we sampled the posterior of the original nonlinear model using \mcmc.
The \mcmc\ calculation used the original model prediction $m(\theta)$ rather than its \dali\ expansion and therefore provided an independent numerical check of the expected equivalence between the original and \dali\ posteriors for this benchmark.
The standard Fisher shift was shown as the leading-order local bias prediction.
Beyond the Gaussian limit, the \map\ and posterior mean are distinguishable and were therefore treated separately.

For a predicted parameter shift $\Delta\theta_{\rm pred}$ and the corresponding numerical reference $\Delta\theta_{\rm \num}$, we quantify the joint difference using the Fisher distance
\begin{equation}
    \label{eq:fisher_distance}
    d_\fish
    =
    \sqrt{
        \left(
        \Delta\theta_{\rm pred}
        -
        \Delta\theta_{\rm \num}
        \right)^{\transp}
        F
        \left(
        \Delta\theta_{\rm pred}
        -
        \Delta\theta_{\rm \num}
        \right)
    }.
\end{equation}
Here $F$ is the Fisher matrix evaluated at the fiducial point, so $d_\fish$ measures the discrepancy between the predicted and numerical shifts relative to the local parameter covariance $F^{-1}$.

For individual parameter components, we additionally define the absolute fractional difference
\begin{equation}
    \epsilon_a =
    \frac{|\Delta\theta_{\rm pred}^a-\Delta\theta_{\rm \num}^a|}
    {|\Delta\theta_{\rm \num}^a|}.
\end{equation}
These fractional differences provide a useful componentwise diagnostic but become ill conditioned when the numerical reference shift approaches zero.
In this regime, a small absolute discrepancy can produce an arbitrarily large fractional difference.
The Fisher distance remains well defined and is therefore used as the primary measure of joint agreement between methods.

The numerical \dali\ posterior mean was evaluated on a regular grid in coordinates $z=L^{\transp}(\theta-\theta_0)$, where $F=LL^{\transp}$ is the Cholesky decomposition of the fiducial Fisher matrix.
The grid was centred on the analytic biased \dali\ \map, with its domain and resolution validated in Appendix~\ref{app:numerical-validation}.

\section{Posterior reconstruction under systematic mismatch}
\label{sec:posterior_reconstruction}

We now examine the accuracy of the Fisher and analytic \dali\ shift predictions for the nonlinear benchmark summarised in Table~\ref{tab:benchmark-settings} and relate their behaviour to the posterior reconstructed under data--model mismatch.
We first connect the \map\ and posterior mean shifts for the combined $\alpha+\beta$ systematic to the corresponding posterior deformation, before examining how the response depends on the data-space direction of the mismatch.

\subsection{Systematic parameter shifts}
\label{subsec:systematic_parameter_shifts}

\begin{figure*}[h!]
    \centering

    \newcommand{\shiftfigwidth}{0.65\textwidth}

    \includegraphics[width=\shiftfigwidth]{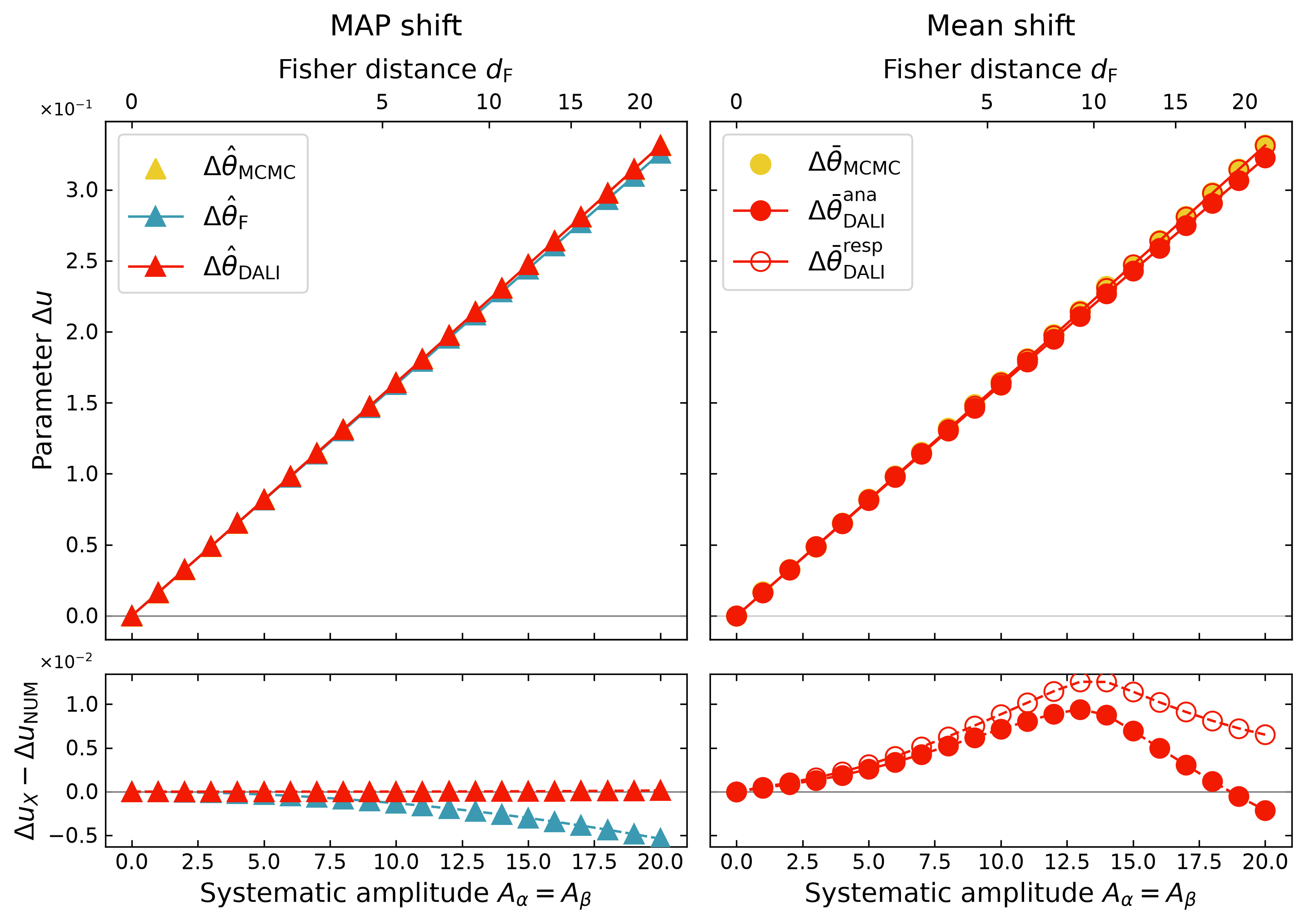}

    \small (a) Parameter $u$.

    \vspace{0.5em}

    \includegraphics[width=\shiftfigwidth]{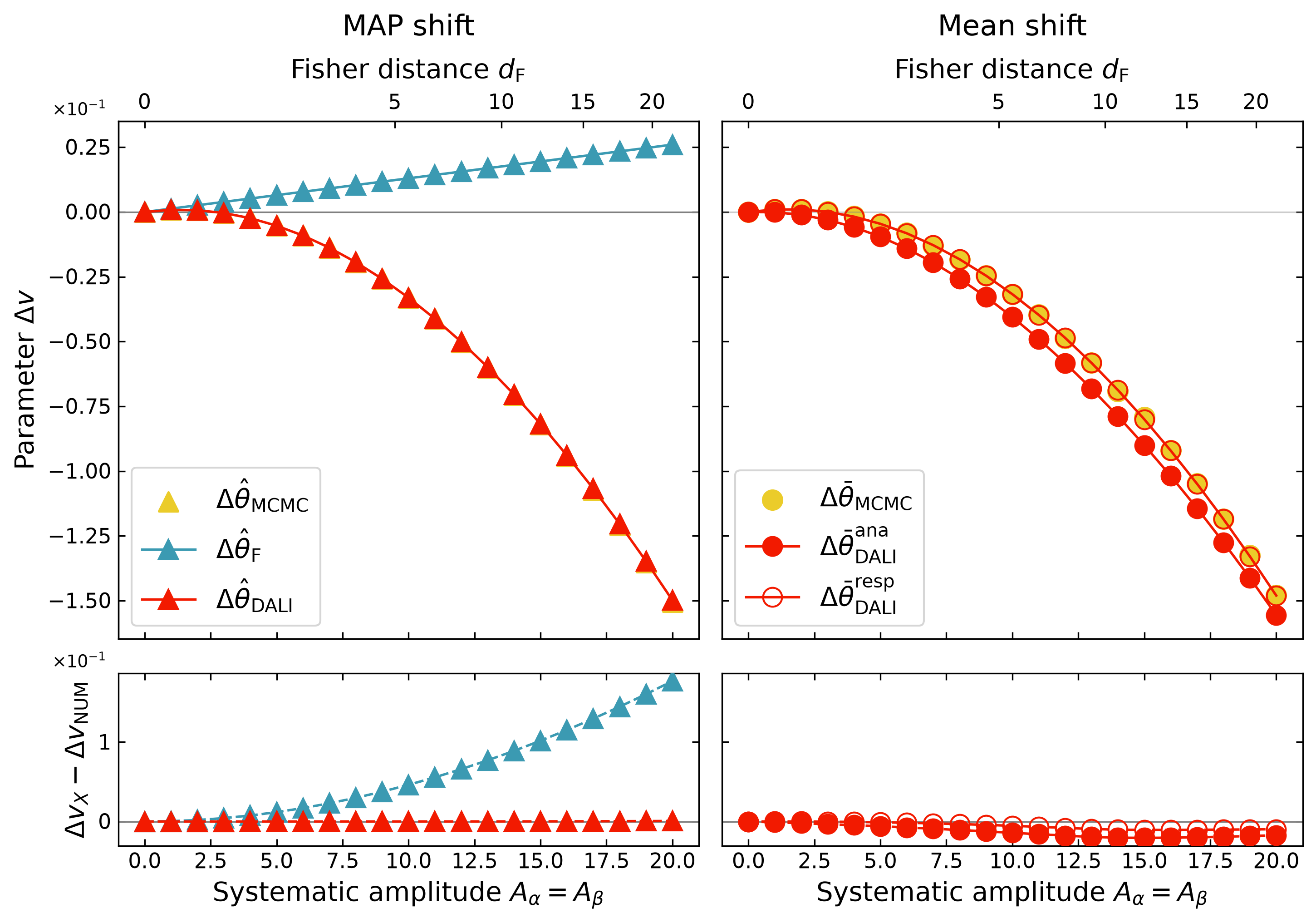}

    \small (b) Parameter $v$.

    \caption{
        Systematic \map\ and posterior mean shifts for the combined $\alpha+\beta$ systematic as a function of systematic amplitude.
        The upper panels compare the Fisher and \dali\ predictions with the \mcmc\ shifts, with the corresponding Fisher distance $d_{\rm F}$ shown on the upper axes.
        The lower panels show the differences between the analytic predictions and the directly evaluated numerical \dali\ \map\ and posterior mean.
    }
    \label{fig:combined-shifts}
\end{figure*}

Figures~\ref{fig:combined-shifts}(a) and \ref{fig:combined-shifts}(b) show the \map\ and posterior mean shifts in $u$ and $v$, respectively, as the common amplitude of the combined $\alpha+\beta$ systematic is increased.
Here we first consider the joint response to the combined data--model mismatch; its relation to the responses of the individual systematics is examined in Section~\ref{subsec:systematic_direction}.
The two parameters exhibit qualitatively different responses to the same data--model mismatch.

The shift in $u$ remains close to linear across the full amplitude range.
Fisher captures the dominant displacement but develops an increasing residual as the data--model mismatch grows, while the analytic \dali\ \map\ prediction closely follows the numerical and \mcmc\ results throughout the scan.

The response in $v$ is strongly nonlinear.
While the Fisher prediction increases linearly in the positive direction, the numerical \map\ instead turns negative and grows rapidly in magnitude.
At $A=20$, the numerical result gives $\Delta\hat{v}=-0.15055$, compared with $\Delta\hat{v}=0.02600$ from Fisher and $\Delta\hat{v}=-0.14985$ from \dali.
The \dali\ prediction therefore captures both the reversal in direction and the subsequent nonlinear growth of the \map\ shift.

The posterior mean exhibits a similarly nonlinear response.
For $u$, the mean shift remains close to linear, whereas for $v$ it turns negative and grows in magnitude as the systematic amplitude increases.
Both the analytic \dali\ mean and the response expansion reproduce these trends and remain close to the numerical mean across the scan.

\subsection{Posterior reconstruction}
\label{subsec:posterior_reconstruction}

\begin{figure}[h!]
    \centering
    \includegraphics[width=0.55\textwidth]{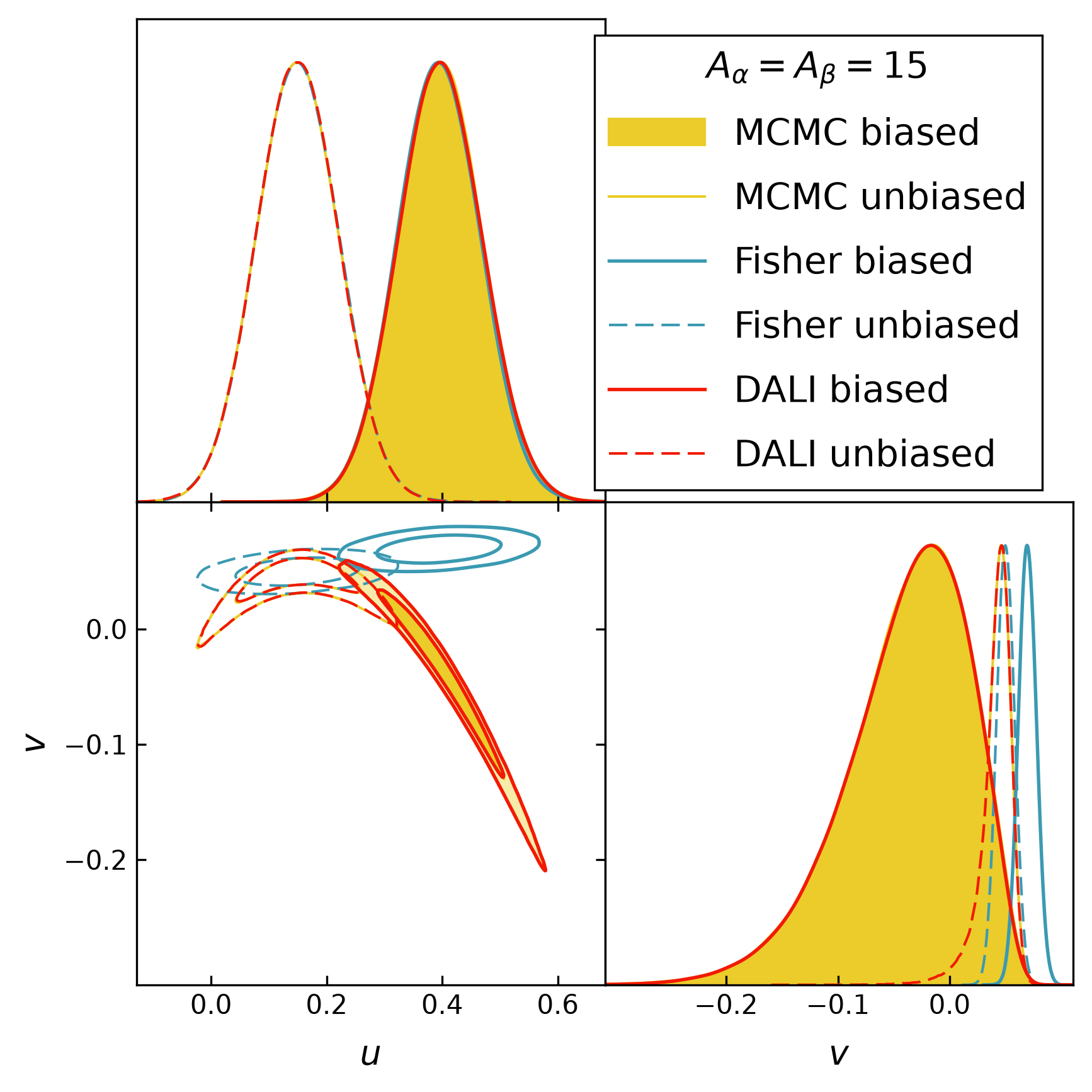}
    \caption{
        Posterior reconstruction under the combined systematic mismatch at $A_\alpha=A_\beta=15$.
        The 1D marginalized posteriors and 2D joint posterior are shown for the \mcmc\ posterior (yellow), Fisher approximation (blue), and \dali\ reconstruction (red).
        For \mcmc, the biased posterior is filled and the unbiased reference is unfilled with solid contours; for Fisher and \dali, the biased posteriors are solid and the unbiased references are dashed, all unfilled.
        The \dali\ reconstruction closely reproduces the systematic shift of the \mcmc-sampled posterior, while Fisher deviates from the nonlinear posterior and, for parameter $v$, misestimates the location of the posterior peak.
    }
    \label{fig:posterior-reconstruction}
\end{figure}

The parameter shifts do not show how the posterior itself deforms under data--model mismatch.
We therefore compare the Fisher and \dali\ posterior reconstructions directly with the full nonlinear \mcmc\ posterior.
Figure~\ref{fig:posterior-reconstruction} shows this comparison for the combined systematic at $A_\alpha=A_\beta=15$, including both the biased and unbiased posteriors.

The nonlinear posterior is already visibly non-Gaussian in the absence of data--model mismatch.
As the mismatch increases, its location and shape both change rather than undergoing a rigid translation.
Fisher captures the local Gaussian displacement but not this evolving curvature, consistent with its increasingly inaccurate shift in $v$.

As established in Section~\ref{subsec:nonlinear_benchmark}, the triplet \dali\ posterior reproduces the full nonlinear posterior for this benchmark up to numerical derivative errors.
Its agreement with the full nonlinear posterior in both the unbiased and biased cases therefore provides a consistency check of the reconstruction.
The substantial motion toward negative $v$ illustrates the posterior deformation whose associated \map\ and mean shifts are captured by the analytic predictions.

\subsection{Direction dependence and nonadditivity}
\label{subsec:systematic_direction}

\begin{figure*}[h!]
    \centering

    \newcommand{\blockwidth}{0.7\textwidth}

    \begin{minipage}{\blockwidth}
        \centering

        \begin{minipage}[t]{0.48\textwidth}
            \centering
            \includegraphics[width=\textwidth]{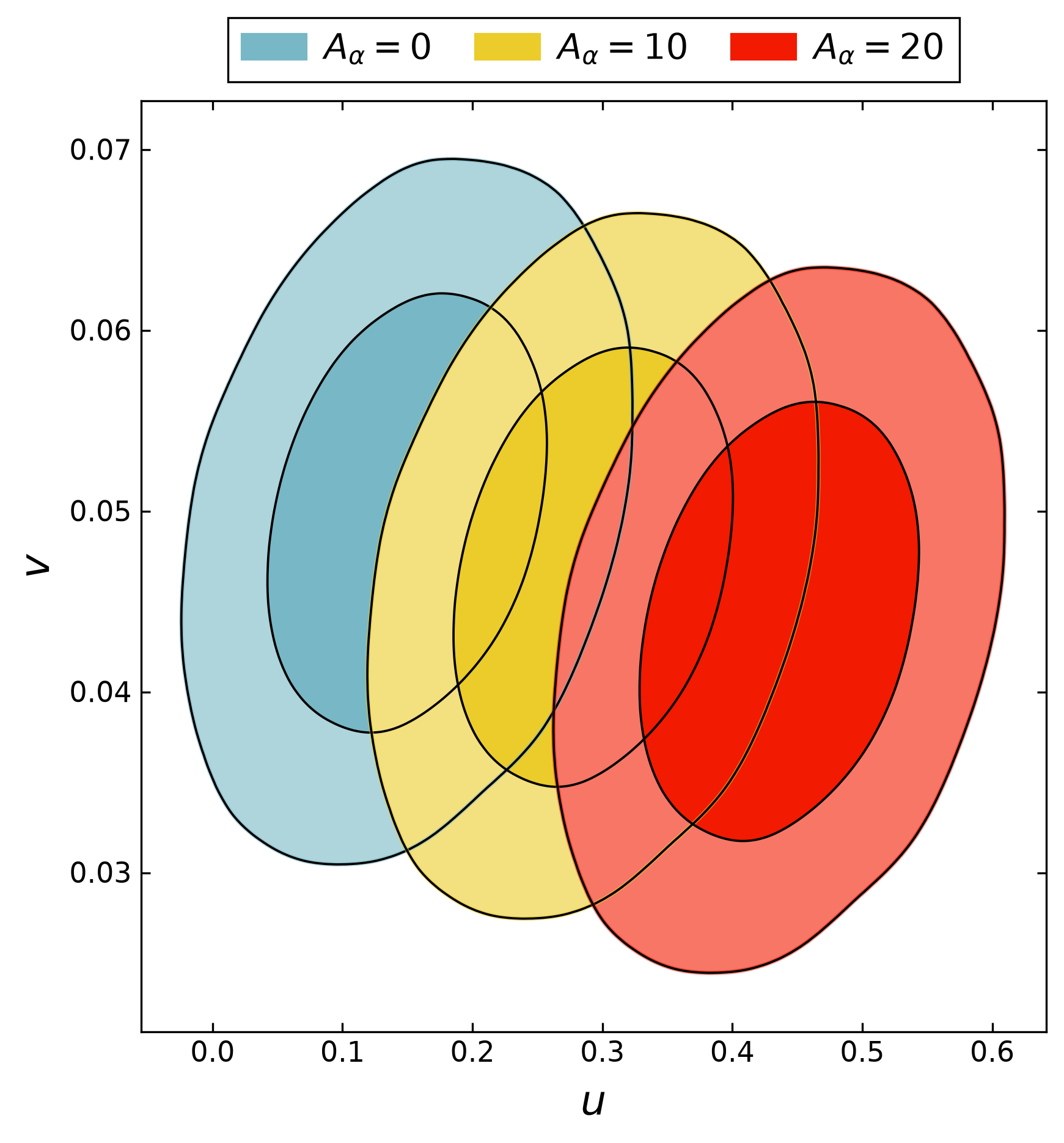}

            \small (a) Fisher, systematic $\alpha$.
        \end{minipage}
        \hfill
        \begin{minipage}[t]{0.48\textwidth}
            \centering
            \includegraphics[width=\textwidth]{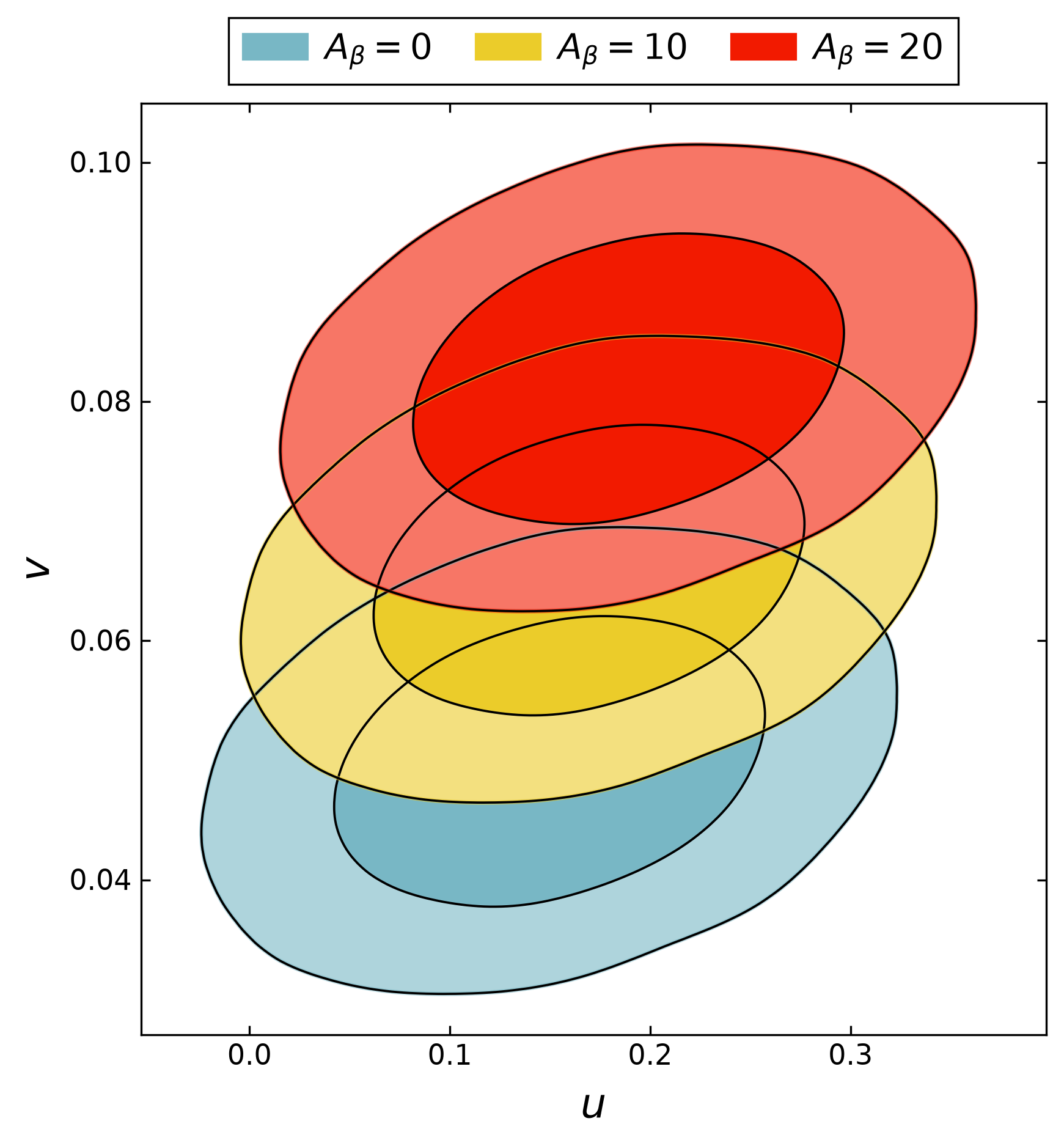}

            \small (b) Fisher, systematic $\beta$.
        \end{minipage}

        \vspace{0.8em}

        \begin{minipage}[t]{0.48\textwidth}
            \centering
            \includegraphics[width=\textwidth]{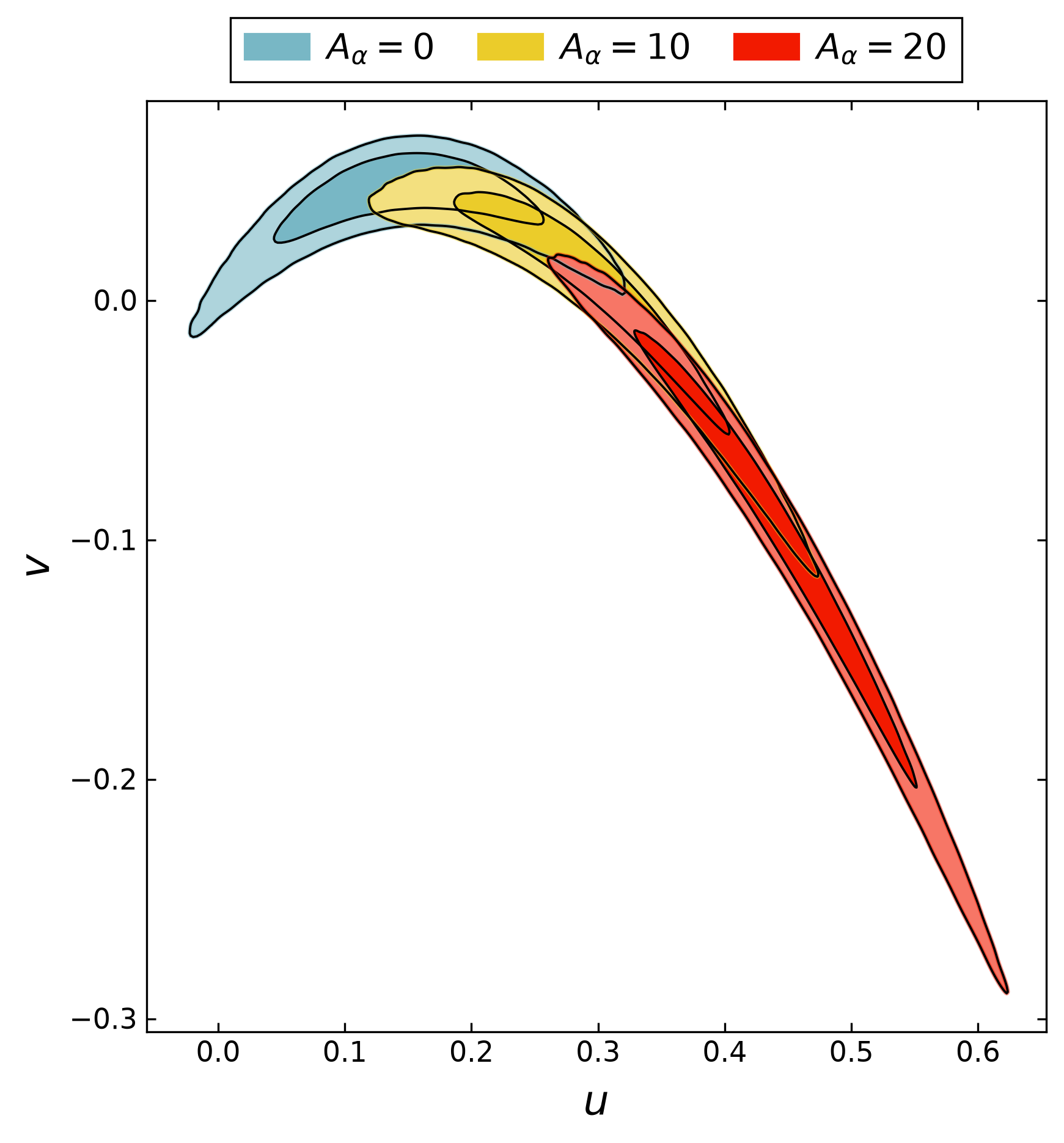}

            \small (c) \dali, systematic $\alpha$.
        \end{minipage}
        \hfill
        \begin{minipage}[t]{0.48\textwidth}
            \centering
            \includegraphics[width=\textwidth]{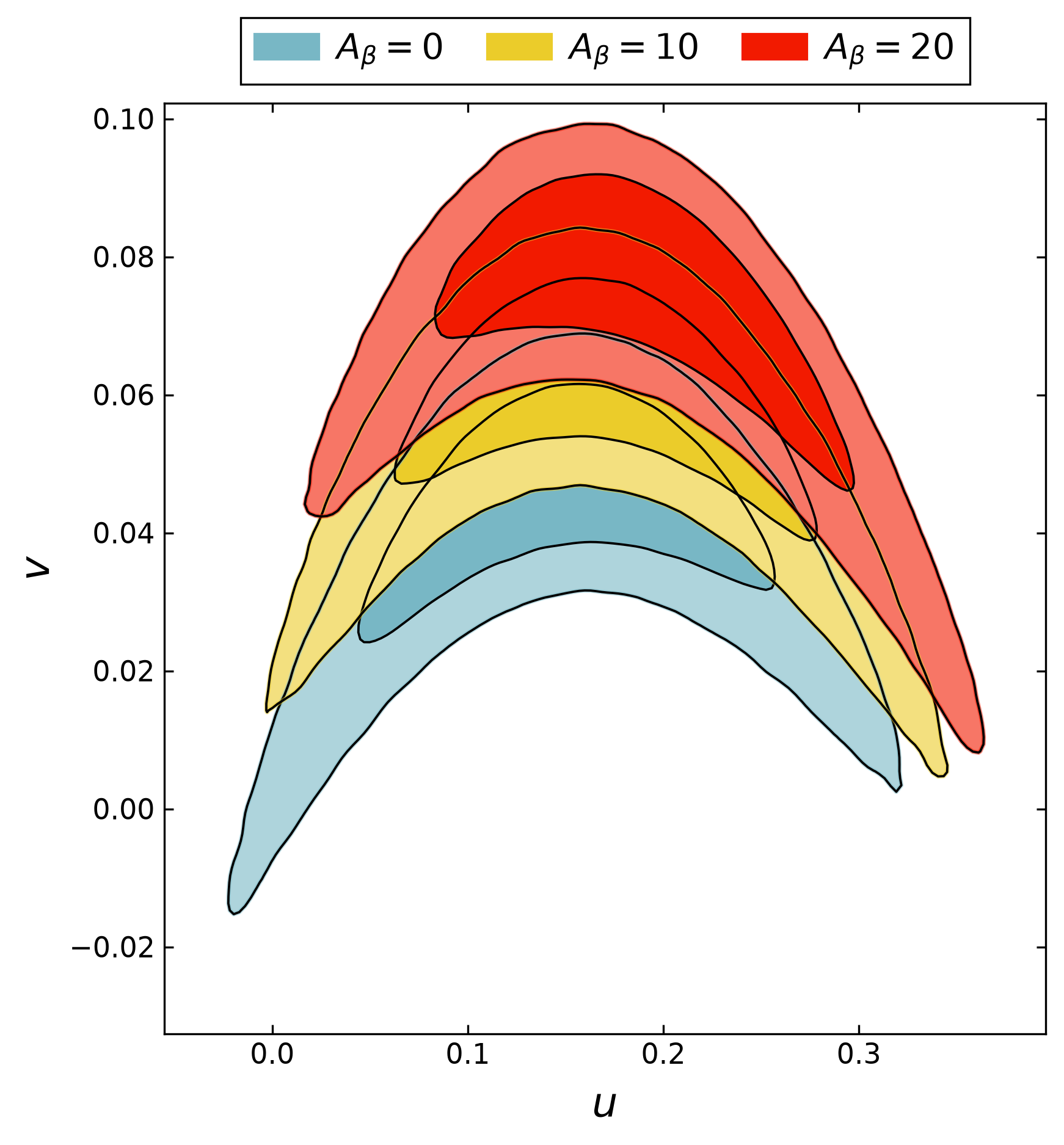}

            \small (d) \dali, systematic $\beta$.
        \end{minipage}
    \end{minipage}

    \caption{
        Evolution of the Fisher (top row) and \dali\ (bottom row) posteriors under the individual systematic mismatch directions $\alpha$ and $\beta$, shown at amplitudes $A=0$, $10$, and $20$.
        The same nominal amplitude produces strongly direction-dependent parameter responses.
        Fisher describes the corresponding local Gaussian displacement, while \dali\ captures the nonlinear displacement and deformation of the posterior as the mismatch increases.
        The contrast between $\alpha$ and $\beta$ demonstrates that the amplitude of the injected mismatch alone does not determine the magnitude or nonlinearity of the resulting parameter response.
    }
    \label{fig:individual-posterior-evolution}
\end{figure*}

The nonlinear parameter response depends not only on the magnitude of the data--model mismatch but also on its direction in data space.
Figure~\ref{fig:individual-posterior-evolution} shows the posterior evolution for the individual $\alpha$ and $\beta$ systematic templates at $A_\alpha, A_\beta=0$, $10$, and $20$, illustrating the distinct nonlinear responses that underlie the combined behaviour.

The two systematic templates produce markedly different posterior responses.
Increasing $A_\alpha$ drives the posterior along the curved parameter degeneracy, producing a large displacement in $u$, an increasingly negative shift in $v$, and substantial deformation by $A_\alpha=20$.
In contrast, increasing $A_\beta$ produces a much milder response: the posterior remains comparatively close to its unbiased location and largely preserves its overall shape over the same amplitude range.

Because the systematic templates have different normalisations, the same amplitude $A$ does not in general correspond to the same magnitude of data--model mismatch for $\alpha$ and $\beta$.
The covariance-weighted distance $\ddata$ provides a common measure with which to compare the severity of different mismatches.
The posterior response can therefore depend on two distinct properties of the systematic perturbation: its magnitude in data space and its direction.

The nonlinear response also implies that the effects of multiple systematics need not combine additively in parameter space.
The Fisher shift in Eq.~\eqref{eq:gaussian_shift} is linear in the injected data mismatch and is therefore additive, whereas the higher-order \dali\ \map\ and posterior-mean responses in Eqs.~\eqref{eq:dalibias_general} and \eqref{eq:mean_bias_main_scheme} contain nonlinear dependence on the mismatch and need not preserve this additivity.
Although the injected templates are exactly additive in data space,
\begin{equation}
    \Delta m_{\alpha,\beta}(A_\alpha,A_\beta)
    =
    \Delta m_\alpha(A_\alpha)
    +
    \Delta m_\beta(A_\beta),
\end{equation}
additivity is not guaranteed for the inferred parameter response.
For the common-amplitude configuration considered here, $A_\alpha=A_\beta=A$, we denote the joint mismatch by $\Delta m_{\alpha+\beta}(A)$ and define the nonadditive contribution as
\begin{equation}
    \Delta\boldsymbol{\theta}_{\rm nonadd}(A)
    =
    \Delta\boldsymbol{\theta}_{\alpha+\beta}(A)
    -
    \left[
        \Delta\boldsymbol{\theta}_{\alpha}(A)
        +
        \Delta\boldsymbol{\theta}_{\beta}(A)
    \right].
\end{equation}
This quantity vanishes when the parameter response is additive.

\begin{figure*}[h!]
    \centering

    \newcommand{\nonaddwidth}{0.70\textwidth}

    \begin{minipage}{\nonaddwidth}
        \centering

        \includegraphics[width=\textwidth]{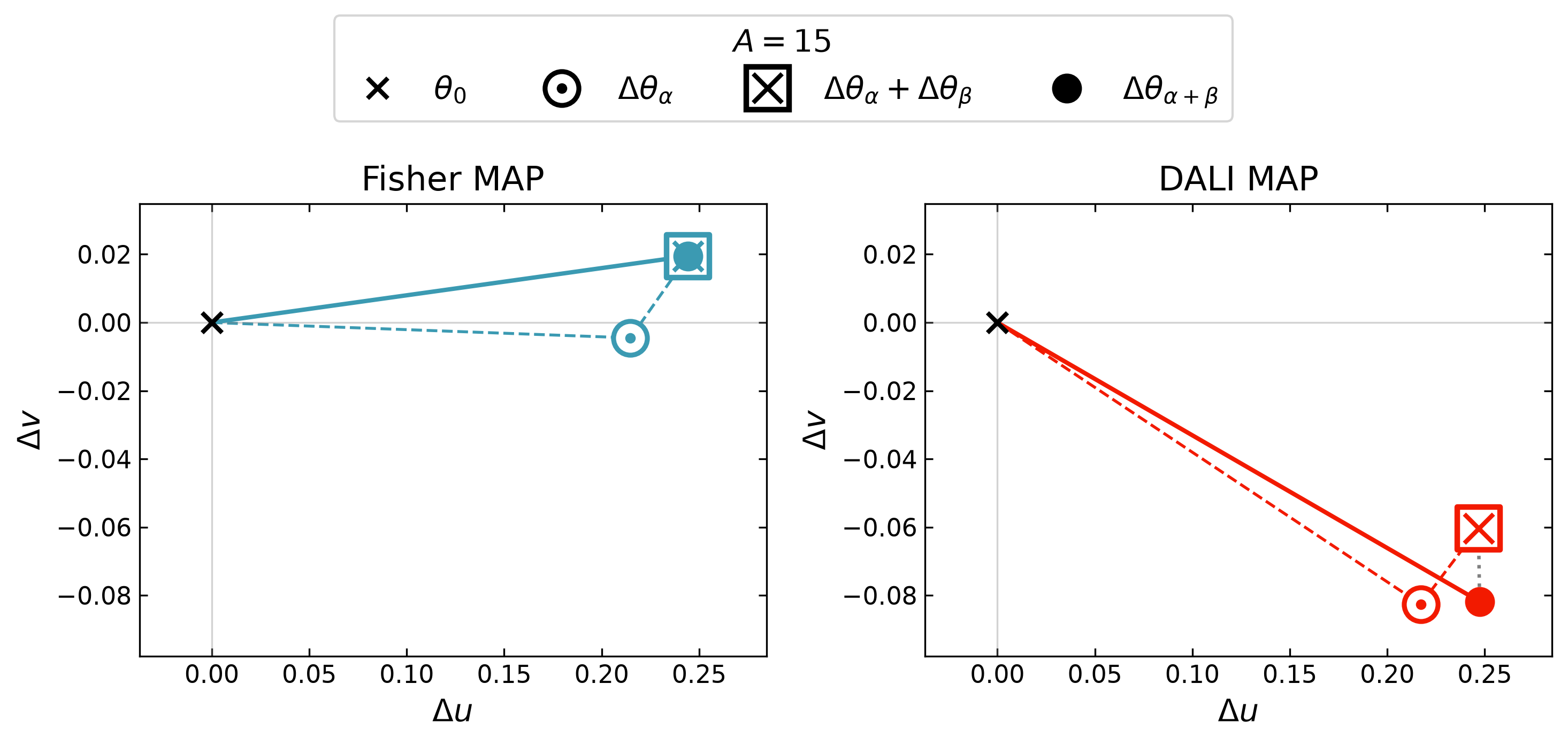}

        \small (a) Parameter-space additivity geometry at $A_\alpha=A_\beta=15$.

        \vspace{0.8em}

        \includegraphics[width=\textwidth]{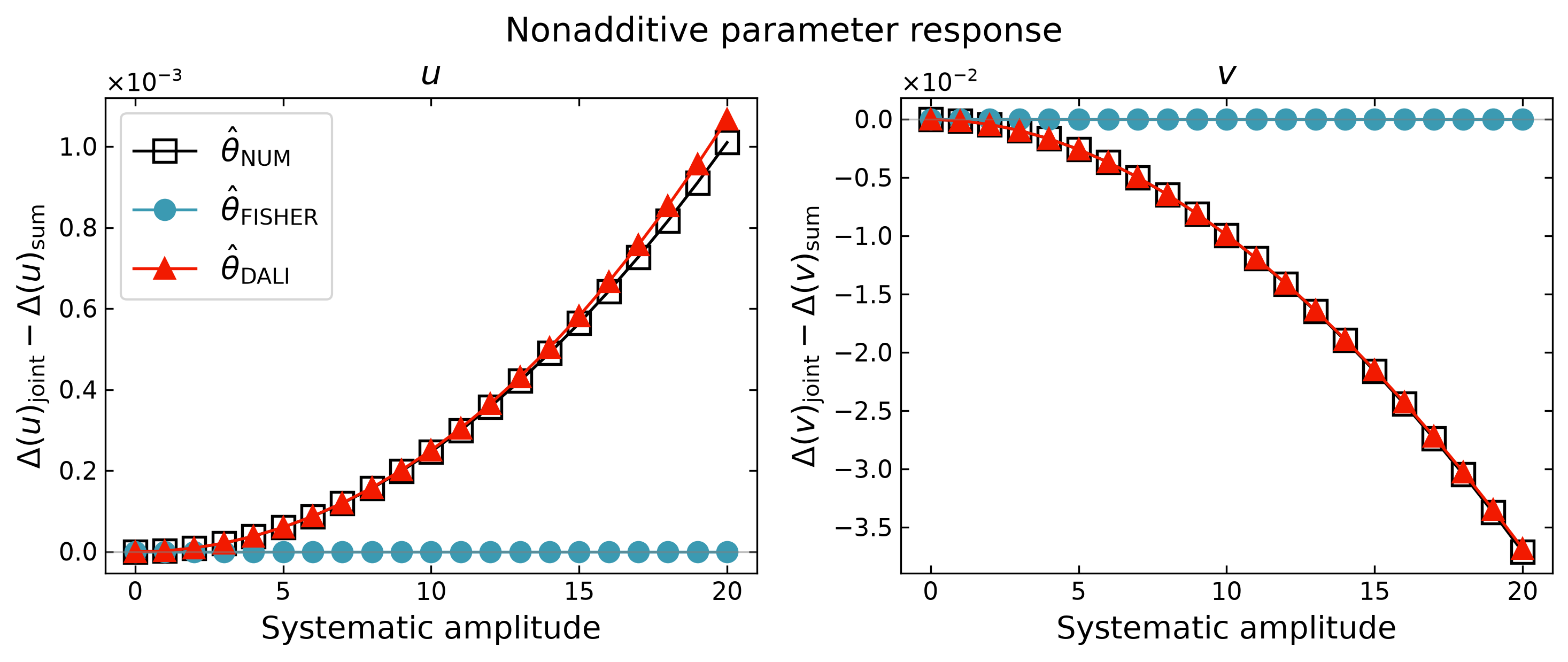}

        \small (b) Nonadditive response as a function of systematic amplitude.
    \end{minipage}

    \caption{
        Nonadditive parameter response for the combined $\alpha+\beta$ systematic.
        Panel (a) compares the additive prediction $\Delta\boldsymbol{\theta}_{\alpha}+\Delta\boldsymbol{\theta}_{\beta}$ with the joint response $\Delta\boldsymbol{\theta}_{\alpha+\beta}$ at $A_\alpha=A_\beta=15$ for the Fisher and \dali\ \map\ predictions.
        Panel (b) shows the corresponding residual over the full amplitude range, including the numerical nonlinear \map\ reference.
        The Fisher response (blue circles) remains additive, while the \dali\ prediction (red triangles) closely follows the nonadditive numerical response (black open squares).
    }
    \label{fig:nonadditivity}
\end{figure*}

Figure~\ref{fig:nonadditivity}(a) illustrates this construction geometrically at $A_\alpha=A_\beta=15$.
For Fisher, the additive prediction and joint response coincide, as expected from the linear dependence of the Fisher shift on the injected data mismatch.
For \dali, the additive prediction and joint response are separated, indicating a nonzero contribution from the nonlinear parameter response.
Figure~\ref{fig:nonadditivity}(b) shows how this residual develops across the full amplitude range.
The Fisher residual remains zero up to numerical precision, whereas the numerical nonlinear \map\ develops a nonzero residual as the mismatch increases.
The analytic \dali\ \map\ prediction closely follows the numerical result across the scan.

Hence, even when systematic contributions are exactly additive in data space, their inferred parameter shifts need not be additive once nonlinear posterior response becomes important.
The nonadditive contribution isolates the part of the parameter response that arises from the joint propagation of multiple additive data-space mismatches through the nonlinear inference map.
Its magnitude and direction therefore characterise the resulting nonadditivity of the parameter-space response as the inference departs from the locally linear regime.

\section{Discussion and summary}
\label{sec:discussion_and_summary}

In this work, we have developed and validated a framework for propagating systematic mismatches between data and model into parameter inference beyond the Fisher bias approximation.
Building on the \dali\ expansion \citep{Sellentin_2014}, we derived separate predictions for the systematic shifts of the posterior \map\ and mean, which coincide in the Gaussian limit but probe distinct properties of a non-Gaussian posterior.
For the \map, the stationarity condition of the posterior leads to a nonlinear generalisation of the standard Fisher bias equation.
For the posterior mean, we derived both a closed analytic approximation and a complementary response expansion in the systematic amplitude that retains the complete non-Gaussian unbiased \dali\ posterior as its reference distribution.

We validated these predictions using a controlled nonlinear benchmark for which the full posterior can be evaluated numerically and independently sampled with \mcmc.
The benchmark probes nonlinear posterior response across a range of systematic amplitudes, with non-Gaussian structure already present in the uncontaminated posterior.
Across the full range considered, extending to a covariance-weighted data-space mismatch of $\ddata=5$, the higher-order predictions capture the nonlinear posterior response substantially better than the Fisher approximation.
In the combined systematic case, the analytic \dali\ \map\ prediction remains accurate even when the Fisher prediction fails to reproduce the direction of the shift in one parameter.
At $\ddata=5$, the Fisher prediction differs from the numerical \dali\ \map\ by $23.1$ in Fisher distance, compared with $0.091$ for the analytic \dali\ prediction.
The closed analytic \dali\ mean captures the nonlinear trend but develops appreciable discrepancies as the mismatch increases, whereas the response expansion remains in close agreement with direct numerical integration across the systematic scan.
At the same mismatch significance, the response expansion differs from the numerical \dali\ mean by only $0.035$ in Fisher distance, with componentwise errors of $0.02\%$ and $0.18\%$.
As discussed in Section~\ref{sec:modelling_and_inference}, the cubic benchmark isolates errors in the analytic shift predictions and truncated response expansion from truncation errors in the \dali\ representation itself.
The Fisher approximation retains only the local Gaussian structure.

These results clarify the role of the Fisher bias approximation.
Fisher captures the leading local linear parameter response to a systematic mismatch.
It does not, however, describe how that response evolves as the posterior is driven away from the expansion point and begins to probe the nonlinear parameter-to-data mapping.
This distinction is particularly clear in the benchmark considered here.
One parameter remains close to its leading linear response over a substantial range of systematic amplitudes, whereas nonlinear effects shift the other to values differing from the Fisher prediction in both magnitude and sign.
The failure is therefore not simply that Fisher underestimates or overestimates the size of a bias.
Nonlinear evolution can change the direction of the inferred parameter shift itself.

The higher-order \map\ expansion gives a direct mathematical description of this departure.
At leading order, the stationarity condition reduces to the standard Fisher solution.
Contractions of higher-order model derivatives with \dali\ tensors then generate nonlinear corrections to the systematic response, including couplings between different systematic directions when multiple systematics are present.
These corrections drive the \map\ response away from the linear Fisher extrapolation.
In this sense, the Fisher shift is the tangent response at the expansion point, while the higher-order terms describe how that response bends as the systematic drives the inferred parameters into nonlinear regions of the model.

The posterior mean provides a complementary view because it depends on the probability mass across the full posterior rather than only on the location of its maximum.
The local analytic \dali\ mean approximation estimates the mean--\map\ offset through perturbative corrections around a Gaussian centred on the \map.
We have obtained the offset relative to the \map\ due to data--model mismatch by subtracting from this the result generated purely by the \dali\ tensors.
Unlike the \map\ calculation, which follows from a local stationarity condition, the mean responds to the shape and asymmetry of the probability distribution over the full region carrying appreciable posterior mass.

The response formulation provides an especially direct interpretation of the systematic mean shift.
As shown in Eq.~\eqref{eq:semi-analytic}, the systematic mean response is governed by expectation values involving the parameters and powers of the mismatch-induced response function $B$, evaluated under the unbiased \dali\ posterior.
At first order, a parameter moves to the extent that its fluctuations in the unbiased posterior are correlated with the systematic perturbation.
Regions of the posterior that are preferentially upweighted by the mismatch pull the mean toward them, while regions that are downweighted contribute in the opposite direction.
A vanishing covariance therefore implies no first-order mean response, while the sign of the covariance determines the initial direction of the shift.

The amplitude range considered here spans from the local-response regime to substantial systematic mismatches.
At $A=20$, the individual $\alpha$ and $\beta$ systematics reach $\ddata=4.49$ and $4.05$, respectively, while their common-amplitude combination reaches $\ddata=5$.
For the combined systematic, the corresponding data-space displacement is $(0.46,0.26)$ relative to the fiducial model prediction $m_0=(1,0.8)$.
The upper end of the scan should therefore be interpreted as a deliberate stress test of the Fisher approximation rather than as a small-perturbation regime.
The magnitude of the injected mismatch alone, however, cannot determine the magnitude or character of the resulting parameter bias.
The individual systematic directions examined here produce considerably different posterior evolution because they probe different nonlinear features of the parameter-to-data mapping.
The $\alpha$ systematic drives the posterior far along its curved degeneracy and produces substantial deformation, whereas the $\beta$ systematic produces a comparatively mild displacement.
Although the systematic vectors are exactly additive in data space, their inferred parameter shifts need not be additive once the parameter response becomes nonlinear.
The systematic response therefore depends not only on the magnitude of the mismatch in data space, but also on its direction and on how that direction probes the nonlinear structure of the model and posterior.
The maximum value $\ddata=5$ characterises the range explored for the mismatch directions considered in this benchmark and should not be interpreted as a general validity threshold, since the nonlinear response depends on both the mismatch direction and the local model geometry.

The present validation is intentionally restricted to a controlled low-dimensional benchmark with parameter-independent covariance.
This setting allows the different approximations to be isolated and tested directly against numerical posterior calculations and independent \mcmc\ sampling, but does not establish the computational scaling or accuracy of the framework in a full high-dimensional cosmological analysis.
Applications to realistic cosmological likelihoods, larger nuisance-parameter spaces, and parameter-dependent covariances therefore remain for future work.

These capabilities are particularly relevant for precision cosmology.
\stagefour\ surveys \citep{lsst_stageiv_ivezic, simonsobs_stageiv, desi_stageiv_abareshi, euclid_stageiv_scaramella, roman_stageiv_spergel} will deliver increasingly precise cosmological measurements that rely on complex astrophysical, observational, and calibration models, precisely where the mapping from systematic data mismatches to inferred parameters need not remain linear.
Fully sampling every combination of systematic assumptions is expensive and, even when feasible, posterior samples alone do not directly identify which aspects of the model geometry generate a particular response.
A derivative-based higher-order description provides both a direct approximation to the posterior response and a means of decomposing and interpreting its origin.

The framework extends Fisher bias forecasting by resolving nonlinear changes in the \map\ and posterior mean under systematic mismatch.
In the benchmark, these corrections recover behaviour that the linear prediction misses, including a reversal in the direction of a parameter shift.
More broadly, this provides a route not only to forecasting biases beyond the Gaussian approximation, but to understanding how systematic modelling choices reshape posterior contours, interact with nonlinear parameter degeneracies, and ultimately propagate into cosmological inference.

\section*{Acknowledgments}

The authors thank Alan Heavens for insightful discussions and helpful comments that improved this work, and Marco Bonici for valuable discussions on the Fisher bias formalism.
N\v{S} is grateful to Michael Troxel for his support and encouragement, and for fostering the intellectual freedom to explore the ideas that led to this work.

N\v{S} is supported in part by the US Department of Energy, Office of Science, \doe\ grant DE-SC0010007.
MvdW is supported in part by the Science and Technology Facilities Council via \lofar-\uk~[ST/V002406/1] and \uksrc\ [ST/X002586] and in part by \nasa\ under award number 80NSSC26K0126.

\paragraph{Software acknowledgments}
This work relied extensively on open-source scientific software, and we are grateful to the developers and maintainers whose work makes research like this possible.
We acknowledge the Python programming language\footnote{\url{https://www.python.org}} as the primary development environment, with core numerical functionality provided by \pkg{NumPy} \citep{numpy} and \pkg{SciPy} \citep{scipy}.
The analysis builds on \pkg{DerivKit} \citep{derivkit} for numerical differentiation, higher-order inference, and parameter bias calculations.
Symbolic manipulations required for the higher-order expansions were performed using \pkg{FORM} \texttt{5.0.1} \citep{form}.
Posterior sampling and post-processing were performed using \pkg{emcee} \citep{emcee} and \pkg{GetDist} \citep{getdist}.
Visualization was performed using \pkg{Matplotlib} \citep{matplotlib} and \pkg{cmasher} \citep{cmasher}, with color palettes inspired by the Wes Anderson collection \citep{hvitfeldt_zissou_palette}.

\paragraph{Software availability}
Scripts developed as part of this work are available under \href{https://doi.org/10.5281/zenodo.23087356}{10.5281/zenodo.23087356}.

\section*{Author Contributions}

N\v{S} conceived the project, co-developed the \dali\ bias formalism, including the posterior mean response formulation, and led the numerical implementation, validation, and analysis.
MvdW co-developed the \dali\ bias formalism, led its theoretical and perturbative development, and carried out the \map\ and posterior mean derivations.
ES contributed to discussions on the \dali\ formalism.
N\v{S} and MvdW wrote and revised the manuscript.

% You should give the same name for your .bbl as your main .tex
% since it is a requirement for posting on ArXiv.
\bibliography{dali_bias}

\newpage

% Our Appendix
\appendix
\section{Derivation of posterior \map\ and mean shifts}
\label{ap:ap1}

This section establishes the framework required to quantify shifts in parameter inference induced by systematic model mismatch.
For pedagogical reviews on statistical techniques in cosmology see \eg\ \cite{Trotta_2008}, \cite{heavens2010statisticaltechniquescosmology}, \cite{Verde_2010}.

For Gaussian distributed data $d$ with model prediction $m(\theta)$ with parameter-independent inverse covariance matrix%
\footnote{
    The determinant of $M$ contributes a constant normalisation factor and has been neglected.
}
$M$, the log-likelihood $\llikelihood$ takes the form
\begin{equation}
    \label{eq:app_log_likelihood}
    \llikelihood(d\vert\theta) = -\sfrac{1}{2} (d - m(\theta))^{\transp} M (d - m(\theta)).
\end{equation}
The parameter fiducial point $\theta_0$ contains the parameter values about which the forecast or perturbative expansion is constructed.
In the absence of systematic model mismatch, the fiducial point is taken to coincide with the maximum a posteriori (\map) estimator $\maptheta$ which, in this case, maximizes the likelihood.
Unmodelled systematics can, instead, induce systematic shifts which shift the maximum of the posterior away from the fiducial point, such that $\maptheta\neq\theta_0$.

The shift away from the fiducial point can be written as $\maptheta = \theta_0 + \delta$.
This leads to \eqref{eq:likelihood} which, for convenience, will be repeated here:
\begin{align}
\llikelihood(d\vert\theta_0 + \delta) =&
        \llikelihood(d\vert\theta_0)
            - \sfrac{1}{2} F_{ab} \delta^a \delta^b
            - \sfrac{1}{2} D^1_{abc} \delta^a \delta^b \delta^c
            - (
                \sfrac{1}{3!} T^1_{abcd}
                + \sfrac{1}{8} D^2_{abcd}
            ) \delta^a \delta^b \delta^c \delta^d
            \nonumber\\&\quad
            - \sfrac{1}{12} T^2_{abcde} \delta^a \delta^b \delta^c \delta^d \delta^e
            - \sfrac{1}{72} T^3_{abcdef} \delta^a \delta^b \delta^c \delta^d \delta^e \delta^f
            \nonumber\\&\quad
            + b_a \delta^a
            + \sfrac{1}{2} b_{ab} \delta^a \delta^b
            + \sfrac{1}{3!} b_{abc} \delta^a \delta^b \delta^c.
    \label{eq:app-likelihood}
\end{align}
Recall that $F$, $D$, and $T$ are the Fisher matrix and \dali\ doublet and triplet, defined in \eqref{eq:dali-definition}, respectively, while the $b$ tensors are defined in \eqref{eq:bias-definition}.

\subsection{Gaussian corrections}
\label{app:gaussian-corrections}

The Gaussian approximation truncates the log-likelihood at the order quadratic in the parameter displacement.
The gradient \eqref{eq:likelihood} then truncates at linear order:
\begin{equation}
    b_a - (F_{ab} - b_{ab}) \delta^b = 0.
\end{equation}
As a result, the shift $\shiftgaussian$ can be solved for exactly as
\begin{equation}
    \shiftgaussian^a = A^{ab} b_b,
    \qquad
    A^{ab} = F^{ac} [(1 - bF^{-1})^{-1}]_c{}^b.
    \label{eq:gaussian_shift_app}
\end{equation}
For sufficiently small $b_{ab}$, this expression can be expanded perturbatively as
\begin{equation}
    \delta_{\textrm G}^a = F^{ab} b_b
                    + F^{ab} F^{cd} b_{bc}b_d
                    + F^{ab} F^{cd} F^{ef} b_{bc}b_{de}b_f
                    + \ldots.
\end{equation}
The first term in this expression is the linear Fisher bias shift \eqref{eq:fisher_bias}.

\subsection{Non-Gaussian corrections}

If the parameter-dependence of the model extends beyond the quadratic order the resulting likelihood in parameter space is non-Gaussian.
For such models, the shift of the \map\ generally does not have a closed-form solution.
However, perturbative expansions can still be used to derive analytic expressions that are in close agreement with numerical or observational results.

In Gaussian models the \map\ coincides with the mean $\meantheta$ of the posterior.
In non-Gaussian models these two quantities are generally distinct.
The shift of the \map\ can be computed by extending the derivation in Sec.~\ref{app:gaussian-corrections}.
The shift of the mean can be computed as a correction to the shift of the \map.

\subsubsection{Maximum a posteriori}
\label{app-sec:map}

The gradient for the log-likelihood for non-Gaussian models can be obtained directly from \eqref{eq:likelihood}.
Setting the result equal to zero then results in the following condition for $\delta$:
\begin{align}
 b_a &= A_{ab} \delta^b
    + \sfrac{1}{2} (
        3 D^1_{abc}
        - b_{abc}
    ) \delta^b \delta^c
    + \sfrac{1}{6} (
        4 T^1_{abcd}
        + 3 D^2_{abcd}
    ) \delta^b \delta^c \delta^d
    \nonumber\\
    &\qquad
    + \sfrac{5}{12} T^2_{abcde} \delta^b \delta^c \delta^d \delta^e
    + \sfrac{1}{12} T^3_{abcdef} \delta^b \delta^c \delta^d \delta^e \delta^f.
    \label{eq:non-gaussian-eq}
\end{align}
The solution $\delta_{\textrm\map}$ to this equation can be decomposed as $\delta_{\textrm\map} = \shiftgaussian + \shiftnongaussian$ with $\shiftgaussian$ given by \eqref{eq:gaussian_shift_app}.
To solve this for $\shiftnongaussian$, introduce the expansion parameter $\lambda = \vert\langle d - m_0\rangle\vert$ such that
\begin{equation}
    \shiftgaussian^a = \sum_{i=1}\lambda^i\shiftgaussian{}_{i}^a,
    \qquad
    \shiftnongaussian^a = \sum_{i=2}\lambda^i\shiftnongaussian{}_{i}^a,
    \qquad
    b_{i_1\ldots i_n} = \lambda u_{i_1\ldots i_n}.
    \label{eq:non-gaussian-constraint}
\end{equation}
Under this expansion, \eqref{eq:non-gaussian-eq} becomes a power series in $\lambda$.
The truncation of \eqref{eq:non-gaussian-eq} requires that this power series be truncated at order 5.
In the resulting polynomial, the coefficient at each order in $\lambda$ must vanish separately.
This recursively determines $\shiftnongaussian{}_i$ in terms of the Gaussian shift coefficients, the \dali\ tensors, and lower-order non-Gaussian corrections.
The resulting solutions including non-Gaussian effects up to the \dali\ triplet are
\begin{align}
    \shiftnongaussian{}_{2}^a &=
        - \sfrac{3}{2} F^{ab} D^1_{bcd} \shiftgaussian{}_1^c \shiftgaussian{}_1^d,
\\
    \shiftnongaussian{}_{3}^a &=
        F^{ab} \Big[
            u_{bc} \shiftnongaussian{}_{2}^c
            + \sfrac{1}{2} u_{bcd} \shiftgaussian{}_{1}^c\shiftgaussian{}_{1}^d
            - 3 D^1_{bcd}\shiftgaussian{}_1^c\left(
                \shiftgaussian{}_2^d
                + \shiftnongaussian{}_2^d
            \right)
            - \sfrac{1}{6}(
                4 T^1_{bcde}
                + 3 D^2_{bcde}
            ) \shiftgaussian{}_1^c\shiftgaussian{}_1^d\shiftgaussian{}_1^e
        \Big],
\\\nonumber
    \shiftnongaussian{}_{4}^a &=
        F^{ab} \Big[
            u_{bc}\shiftnongaussian{}_3^c
            + u_{bcd} \shiftgaussian{}_1^c (
                \shiftgaussian{}_2^d
                + \shiftnongaussian{}_2^d
            )
\\&\qquad\nonumber
            - \sfrac{3}{2} D^1_{bcd} (
                2 \shiftgaussian{}_1^c(
                    \shiftgaussian{}_3^d
                    + \shiftnongaussian{}_3^d
                )
                + (
                    \shiftgaussian{}_2^c
                    + \shiftnongaussian{}_2^c
                ) (
                    \shiftgaussian{}_2^d
                    + \shiftnongaussian{}_2^d
                )
            )
\\&\qquad
            - \sfrac{1}{2} (
                4 T^1_{bcde} + 3 D^2_{bcde}
            ) \shiftgaussian{}_1^c\shiftgaussian{}_1^d (
                \shiftgaussian{}_2^e
                + \shiftnongaussian{}_2^e
            )
            - \sfrac{5}{12} T^2_{bcdef} \shiftgaussian{}_1^c\shiftgaussian{}_1^d\shiftgaussian{}_1^e\shiftgaussian{}_1^f
        \Big],
\\
    \shiftnongaussian{}_{5}^a &=
        F^{ab} \Big[
            u_{bc} \shiftnongaussian{}_4^c
            + \sfrac{1}{2} u_{bcd} \left(
                2 \shiftgaussian{}_1^c (
                    \shiftgaussian{}_3^d
                    + \shiftnongaussian{}_3^d
                )
                + (
                    \shiftgaussian{}_2^c
                    + \shiftnongaussian{}_2^c
                ) (
                    \shiftgaussian{}_2^d
                    + \shiftnongaussian{}_2^d
                )
            \right)
\nonumber\\&\qquad
            - 3 D^1_{bcd} \left(
                \shiftgaussian{}_1^c (
                    \shiftgaussian{}_4^d
                    + \shiftnongaussian{}_4^d
                )
                + (
                    \shiftgaussian{}_2^c
                    + \shiftnongaussian{}_2^c
                ) (
                    \shiftgaussian{}_3^d
                    + \shiftnongaussian{}_3^d
                )
            \right)
\nonumber\\&\qquad
            - \sfrac{1}{2} (
                4 T^1_{bcde}
                + 3 D^2_{bcde}
            ) \shiftgaussian{}_1^c (
                \shiftgaussian{}_1^d (
                    \shiftgaussian{}_3^e
                    + \shiftnongaussian{}_3^e
                ) +
                (
                    \shiftgaussian{}_2^d
                    + \shiftnongaussian{}_2^d
                ) (
                    \shiftgaussian{}_2^e
                    + \shiftnongaussian{}_2^e
                )
            )
\nonumber\\&\qquad
            - \sfrac{5}{3} T^2_{bcdef} \shiftgaussian{}_1^c \shiftgaussian{}_1^d \shiftgaussian{}_1^e (
                \shiftgaussian{}_2^f
                + \shiftnongaussian{}_2^f
            )
            - \sfrac{1}{12} T^3_{bcdefg} \shiftgaussian{}_1^c \shiftgaussian{}_1^d \shiftgaussian{}_1^e \shiftgaussian{}_1^f \shiftgaussian{}_1^g
        \Big].
\end{align}

\subsubsection{Mean}
\label{app-sec:mean}

One systematic way of computing the posterior mean $\meantheta$ is to introduce the generating function $Z(t)$, in terms of which the mean can be written as
\begin{equation}
    \meantheta^a = \left.\frac{\partial Z}{\partial t_a}\right\vert_{t=0},
    \qquad
    Z(t) = \log\left[
        \int\exp\left(
            t_i \theta^i + \llikelihood(d \vert \theta)
        \right)\,\dd^{N_{\textrm par}}\theta.
    \right].
    \label{eq:mean-generating-function}
\end{equation}
Note that $N_{\textrm par}$ indicates the number of parameters.

The generating function can, in general, only be integrated exactly for Gaussian likelihoods.
However, one can derive perturbative expansions for it.
We compute two different expressions for $\meantheta$.
In Section \ref{app-sec:mean-response} we express the mean as a formal correction relative to the unbiased mean, which is perturbative in data--model mismatch but exact in \dali.
In the remainder of this section, we express the mean as a correction to the \map, which is exact in data--model mismatch but perturbative in \dali.

The integral in $Z$ can be computed perturbatively by shifting the \map\ \eqref{eq:non-gaussian-constraint} to the origin.
The log-likelihood is then expanded about $\maptheta$, truncating at the level of the \dali\ triplet.
In this case, the log-likelihood can be parametrized as
\begin{equation}
    \label{eqref:log-likelihood-parametrisation}
    \llikelihood(d|\maptheta + \delta) =
            \llikelihood(d|\maptheta)
        - \sum_{n=2}^6\sfrac{1}{n!}C^n_{i_1\dots i_n}\delta^{i_1}\dots\delta^{i_n}.
\end{equation}
The zeroth-order term is an additive constant in the log-likelihood and therefore cancels in \eqref{eq:mean-generating-function}.
The coefficients $-C^n$ can be determined by differentiating \eqref{eq:likelihood} $n$ times with respect to $\delta$ and evaluating the result at the \map.
The result is
\begin{align}
    \label{eq:non-gaussian-1}
    C^n_{i_1\ldots i_n} &= \sum_{r=n}^6\sfrac{1}{(r-n)!}c^r_{i_1\ldots i_n\ldots i_r}\delta_{\textrm \map}^{i_{n+1}}\ldots\delta_{\textrm \map}^{i_r},\\
    c^2_{ab} &= F_{ab} - b_{ab}, \\
    c^3_{abc} &= 3 D^1_{abc} - b_{abc}, \\
    c^4_{abcd} &= 4 T^1_{abcd} + 3 D^2_{abcd}, \\
    c^5_{abcde} &= 10 T^2_{abcde}, \\
    c^6_{abcdef} &= 10 T^3_{abcdef}.
    \label{eq:non-gaussian-2}
\end{align}
Note that the bias and \dali\ tensors are evaluated at the original fiducial point.

To compute the integral in \eqref{eq:mean-generating-function}, expand the non-Gaussian parts of the exponentials and integrate the result term by term.
This can be extended arbitrarily, but is here restricted to the order appropriate to the \dali\ triplet.
The total shift at this order is
\begin{align}
    \label{eq:app-mean-shift}
    \meantheta^a
    &=
    \maptheta^a
    - \sfrac{1}{2} C^3_{ijk} (C^2)^{ia}(C^2)^{jk}
    - \sfrac{1}{8} C^5_{ijklm}(C^2)^{ia}(C^2)^{jk}(C^2)^{lm},
\end{align}
with the $C$ coefficients defined in \eqref{eq:non-gaussian-1}--\eqref{eq:non-gaussian-2}.
The coefficients $(C^2)^{ij}$ are obtained from the inverse of $C^2_{ij}$.

Note that this shift includes both terms arising from data--model mismatch and terms which are due to the non-Gaussian nature of the model.
In order to obtain the genuine shift due to data--model mismatch, one has to subtract from this the value of \eqref{eq:app-mean-shift} evaluated at $\Delta m = 0$.

\subsection{The mean as a formal expansion in data--model mismatch}
\label{app-sec:mean-response}

We now derive a semi-analytic approximation for the response of the posterior mean $\meantheta$ to a data--model mismatch.
In contrast to the preceding expansion about the \map, the complete non-Gaussian \dali\ posterior at zero mismatch is retained as the reference distribution.
The expansion is performed only in the data--model mismatch.

We therefore expand \eqref{eq:mean-generating-function} about the unbiased \dali\ posterior:
\begin{equation}
    \label{eq:app-llikelihood-bias-response}
    \llikelihood(\theta, \Delta m)
    =
    \llikelihood(\theta; 0)
    +
    B(\theta),
\end{equation}
where now%
\footnote{
    Note that, in this context, $B$ is a scalar function, which differs from the matrix notation used in \eqref{eq:gaussian_shift}.
}
\begin{equation}
    \label{eq:app-bias-factor}
    B(\theta)
    =
    \sum_n \sfrac{1}{n!} b_{a_1 \ldots a_n} \theta^{a_1} \dots \theta^{a_n}
\end{equation}
and the $b$ tensors are defined as in \eqref{eq:bias-definition}.

After substituting \eqref{eq:app-llikelihood-bias-response} and \eqref{eq:app-bias-factor} into \eqref{eq:mean-generating-function}, the factor $\exp(B)$ is expanded and the resulting integrals are evaluated term by term.
Terms of like order in $B$ are then collected.
This expansion can be carried to arbitrary order; here, we truncate at cubic order in the data--model mismatch.
The result can then be expressed as a shift relative to the mean $\meantheta^a$ of the unbiased \dali\ posterior:
\begin{align}
\label{eq:app-semi-analytic}
\meantheta^a &= \meantheta^a_0
            + \avg{(\theta^a - \meantheta^a_0)(B-\avg{B})}
            \nonumber\\&\qquad
            + \sfrac{1}{2} \avg{(\theta^a - \meantheta^a_0)(B - \avg{B})^2}
            \nonumber\\&\qquad
            + \sfrac{1}{3!} \avg{(\theta^a - \meantheta^a_0)(B - \avg{B})^3}
            \nonumber\\&\qquad
            - \sfrac{1}{2} \avg{(\theta^a - \meantheta^a_0)(B - \avg{B})}\avg{(B -\avg{B})^2}.
\end{align}
The expectation values on the right-hand side are taken with respect to the full \dali\ posterior $\llikelihood(\theta; 0)$.
While these expectation values are determined numerically, \eqref{eq:app-semi-analytic} expresses the posterior mean response as a polynomial in the data--model mismatch, with the corresponding systematic shift obtained by subtracting $\meantheta_0^a$.

\section{Numerical validation}
\label{app:numerical-validation}

We validated the benchmark calculations at each stage of the analysis.
First, we tested the stability of the analytic \dali\ predictions under changes to the numerical derivative settings.
We then compared the normalised Fisher and \dali\ posteriors with the full nonlinear posterior and tested sensitivity to the adopted parameter bounds.
Finally, we checked convergence of the local \dali\ posterior integration and estimated the Monte Carlo precision of the full nonlinear posterior mean.
Parameter-space displacements were measured using the fiducial Fisher distance defined in Eq.~\eqref{eq:fisher_distance}.

\subsection{Derivative stability}

The analytic \dali\ predictions used throughout the analysis depended on numerical derivatives of the model, evaluated with the local-polynomial estimator \citep{Fornberg_2025} implemented in \derivkit.
Locally, each model component was approximated as
\begin{equation}
    p(x)
    =
    \sum_{k=0}^{d}
    a_k (x-x_0)^k,
\end{equation}
where $x_0$ is the evaluation point, $d$ is the polynomial degree, and $a_k$ are the fitted coefficients; derivatives were obtained analytically from the fitted polynomial.
Further details of the estimator and its numerical implementation are given in the \derivkit\ documentation and the corresponding paper \citep{derivkit}.

We tested whether the resulting bias predictions were sensitive to the polynomial degree.
For each systematic case and amplitude $A=0,\ldots,20$, we computed the analytic \dali\ \map\ displacement, the analytic posterior-mean displacement, and the posterior-mean displacement from the response expansion.
The reference calculation used the default \derivkit\ configuration, with symmetric relative sampling offsets $(0.01,0.02,0.04,0.08)$ and polynomial degree $d=\min[\max(n+2,3),7]$ for a derivative of order $n$.
We compared this reference with fixed polynomial degrees $d=3,4,5,6,7$.

\begin{table*}[h!]
    \centering
    \caption{Numerical derivative stability of the analytic bias predictions.
    Derivative sensitivity is measured relative to the default derivative configuration, while degree convergence is measured between successive local polynomial degrees.
    All entries are Fisher distances.}
    \label{tab:derivative-stability}
    \begin{tabular}{lrrrrrrrr}
        \toprule
        & \multicolumn{4}{c}{Derivative sensitivity}
        & \multicolumn{4}{c}{Degree convergence} \\
        \cmidrule(lr){2-5}
        \cmidrule(lr){6-9}
        Systematic
        & $\Delta\hat{\theta}$
        & $\Delta\bar{\theta}^{\rm ana}$
        & $\Delta\bar{\theta}^{\rm resp}$
        & Max.
        & $\Delta\hat{\theta}$
        & $\Delta\bar{\theta}^{\rm ana}$
        & $\Delta\bar{\theta}^{\rm resp}$
        & Max. \\
        \midrule
        $\alpha$ & $1.78\times10^{-10}$ & $1.51\times10^{-8}$ & $1.28\times10^{-8}$ & $1.51\times10^{-8}$ & $1.86\times10^{-10}$ & $2.21\times10^{-8}$ & $2.14\times10^{-8}$ & $2.21\times10^{-8}$ \\
        $\beta$ & $1.14\times10^{-11}$ & $1.63\times10^{-10}$ & $8.51\times10^{-11}$ & $1.63\times10^{-10}$ & $8.26\times10^{-12}$ & $1.14\times10^{-10}$ & $1.25\times10^{-10}$ & $1.25\times10^{-10}$ \\
        $\alpha+\beta$ & $3.14\times10^{-10}$ & $1.59\times10^{-8}$ & $1.92\times10^{-8}$ & $1.92\times10^{-8}$ & $2.75\times10^{-10}$ & $2.44\times10^{-8}$ & $1.56\times10^{-8}$ & $2.44\times10^{-8}$ \\
        \bottomrule
    \end{tabular}
\end{table*}

For a predicted displacement $\Delta\theta$, derivative sensitivity was quantified by the Fisher distance
\begin{equation}
    \delta_{\mathrm{ref}}
    =
    d_\fish\!\left(
        \Delta\theta_{d}
        -
        \Delta\theta_{\mathrm{ref}}
    \right),
\end{equation}
while sensitivity to successive polynomial degrees was measured by
\begin{equation}
    \delta_{d}
    =
    d_\fish\!\left(
        \Delta\theta_{d}
        -
        \Delta\theta_{d-1}
    \right).
\end{equation}
We performed both tests for the $\alpha$ and $\beta$ systematic directions individually and for their common-amplitude combination, as defined in Table~\ref{tab:benchmark-settings}.

The results are summarised in Table~\ref{tab:derivative-stability}, where each entry gives the maximum Fisher distance over the full amplitude range and tested polynomial degrees.
The analytic \map\ prediction was particularly insensitive to the derivative configuration, with maximum variations of order $10^{-10}$ or smaller.
The posterior-mean predictions were similarly stable.
Across all systematic cases and all three predicted quantities, the largest change relative to the default derivative configuration was $1.92\times10^{-8}$ in Fisher distance, while the largest change between successive polynomial degrees was $2.44\times10^{-8}$.
Both were negligible on the parameter scales relevant to the analysis.

These variations were many orders of magnitude below the parameter displacements relevant to the analysis.
The analytic \map, analytic posterior mean, and response-based posterior mean were therefore numerically insensitive to the tested local-polynomial degree over the full systematic-amplitude range considered here.

\subsection{Posterior validity and parameter bounds}
\label{subsec:posterior_validity_and_bounds}

We compared the Fisher and triplet \dali\ posteriors with the full nonlinear benchmark posterior on the adopted $[-2,2]^2$ domain.
We applied the same uniform prior to all three and normalised each posterior by its own integral over that domain.
This allowed us to compare their full shapes as well as their means.
Across $A=0,\ldots,20$, the largest \dali\ mean difference was $8.17\times10^{-8}$ exact posterior standard deviations, and its largest squared Hellinger distance \citep{Cramer1946_hellinger_distance} $H^2$ was $7.99\times10^{-15}$.
The triplet \dali\ posterior thus agreed with the exact posterior to numerical precision in this benchmark.
By comparison, the Fisher mean difference reached $2.55$ exact posterior standard deviations and its squared Hellinger distance reached $0.998$.
These results are summarised in Table~\ref{tab:posterior-validity-nonlinear-benchmark-full}.

We also compared the exact posterior on $[-2,2]^2$ with the posterior on $[-12,12]^2$ to test whether the adopted bounds affected the result.
The largest mean change was $7.29\times10^{-14}$ outer posterior standard deviations, and the posterior mass outside the adopted $[-2,2]^2$ bounds was below $1.13\times10^{-13}$.
Both were below the numerical resolution of the bounds comparison, so no sensitivity to the adopted bounds was resolved.
Table~\ref{tab:bounds-nonlinear-benchmark-full} gives the results and quadrature checks.
For both comparisons, the log density was quadratic in $v$ at fixed $u$, allowing the $v$ integral to be evaluated analytically, while the remaining integral over $u$ was evaluated by Gauss--Legendre quadrature using 600 and 1200 nodes.

\begin{table*}[t]
    \centering
    \caption{Fisher and \text{\scshape dali} posterior differences from the exact nonlinear benchmark, maximized over $A=0,\ldots,20$.
    Mean differences are in exact posterior standard deviations; $H^2$ compares normalized posteriors on $[-2,2]^2$.
    $Q_Z$ and $Q_\mu$ measure quadrature convergence with 600 and 1200 nodes in $u$, integrating $v$ analytically.}
    \label{tab:posterior-validity-nonlinear-benchmark-full}
    \begin{tabular}{lrrrrrr}
        \toprule
        Systematic & $|\bar{\theta}_{\rm F}-\bar{\theta}_{\rm exact}|/\sigma$ & $|\bar{\theta}_{\text{\scshape dali}}-\bar{\theta}_{\rm exact}|/\sigma$ & Fisher $H^2$ & \text{\scshape dali} $H^2$ & $Q_Z$ & $Q_\mu$ \\
        \midrule
        $\alpha$ & $2.25\times10^{0}$ & $7.58\times10^{-8}$ & $9.92\times10^{-1}$ & $7.99\times10^{-15}$ & $1.06\times10^{-13}$ & $5.92\times10^{-14}$ \\
        $\beta$ & $7.18\times10^{-1}$ & $9.82\times10^{-9}$ & $2.53\times10^{-1}$ & $8.88\times10^{-16}$ & $8.35\times10^{-14}$ & $9.33\times10^{-15}$ \\
        $\alpha+\beta$ & $2.55\times10^{0}$ & $8.17\times10^{-8}$ & $9.98\times10^{-1}$ & $4.44\times10^{-15}$ & $1.46\times10^{-13}$ & $7.57\times10^{-14}$ \\
        \bottomrule
    \end{tabular}
\end{table*}

\begin{table*}[t]
    \centering
    \caption{Sensitivity of the exact nonlinear benchmark posterior to parameter bounds, maximized over $A=0,\ldots,20$.
    $Z_0/Z_R$ is the likelihood integral ratio for the adopted $[-2,2]\times[-2,2]$ and outer $[-12,12]\times[-12,12]$ domains.
    Mean and standard deviation differences are scaled by the outer posterior standard deviation.
    Quadrature checks compare 600 and 1200 nodes in $u$; $v$ is integrated analytically.}
    \label{tab:bounds-nonlinear-benchmark-full}
    \begin{tabular}{lrrrrrr}
        \toprule
        Systematic & $|1-Z_0/Z_R|$ & $Q_r$ & $Q_Z$ & $\max|\bar{\theta}_0-\bar{\theta}_R|/\sigma_R$ & $\max|\sigma_0-\sigma_R|/\sigma_R$ & $Q_\mu$ \\
        \midrule
        $\alpha$ & $1.13\times10^{-13}$ & $1.80\times10^{-11}$ & $1.79\times10^{-11}$ & $5.31\times10^{-14}$ & $5.40\times10^{-14}$ & $5.32\times10^{-10}$ \\
        $\beta$ & $1.11\times10^{-14}$ & $2.17\times10^{-11}$ & $2.16\times10^{-11}$ & $1.52\times10^{-14}$ & $1.19\times10^{-14}$ & $6.49\times10^{-10}$ \\
        $\alpha+\beta$ & $9.99\times10^{-14}$ & $1.99\times10^{-11}$ & $1.98\times10^{-11}$ & $7.29\times10^{-14}$ & $6.01\times10^{-14}$ & $5.92\times10^{-10}$ \\
        \bottomrule
    \end{tabular}
\end{table*}

\subsection{Local \dali\ posterior}

The biased \dali\ posterior was evaluated on a two-dimensional regular grid in Fisher-whitened parameter coordinates.
The integration domain spanned $[-R,R]$ along each whitened parameter direction and was centred on the analytic biased \dali\ \map.

We determined the integration settings from separate domain and grid-resolution scans over $A=0,\ldots,20$, including both individual systematic directions and their common-amplitude combination.
For the domain scan, successive posterior means were required to satisfy
\begin{equation}
    \Delta_R
    =
    \left\|
        \bar{\theta}(R_{i+1})
        -
        \bar{\theta}(R_i)
    \right\|_\fish
    <
    0.01
\end{equation}
for three consecutive increases in $R$.
At the selected radius, the grid spacing was determined analogously from
\begin{equation}
    \Delta_h
    =
    \left\|
        \bar{\theta}(h_{i+1})
        -
        \bar{\theta}(h_i)
    \right\|_\fish
    <
    0.01
\end{equation}
for three consecutive refinements.
The most conservative settings required across all tested cases were adopted.
The resulting integration prescription was $R=46$, $h=0.04$, and $N=2301$.
The limiting domain was set by the combined systematic case at the largest tested amplitudes.

Using these independently selected settings, we verified the numerical stability of the local \dali\ posterior by comparing the $N=2301$ grid with a refined $N=4601$ grid at fixed $R=46$.
For the posterior mean, we required
\begin{equation}
    d_\fish\!\left(
        \meantheta_{4601}
        -
        \meantheta_{2301}
    \right)
    <
    0.01.
\end{equation}

The numerical \map\ was localized only to the finite grid resolution.
We therefore compared its displacement with the combined half-cell scale,
\begin{equation}
    \epsilon_{\map}(N_1,N_2)
    =
    \frac{1}{\sqrt{2}}
    \sqrt{
        h(N_1)^2
        +
        h(N_2)^2
    },
    \qquad
    h(N)
    =
    \frac{2R}{N-1},
\end{equation}
and required
\begin{equation}
    d_\fish\!\left(
        \maptheta_{4601}
        -
        \maptheta_{2301}
    \right)
    \leq
    \epsilon_{\map}.
\end{equation}

\subsection{\mcmc\ sampling convergence}

The full nonlinear posterior was sampled using the \emcee\ ensemble sampler \citep{emcee} (version 3.1.6) with a fixed random seed of $96$.
The chains used $N_{\rm walk}=32$ walkers with $60,000$ steps per walker, of which the first $6000$ were discarded as burn-in, with no additional thinning.
For each parameter, we estimated the effective number of independent posterior samples from the integrated autocorrelation time $\tau$ as
\begin{equation}
    N_{\rm eff}
    =
    \frac{N_{\rm walk}N_{\rm post}}{\tau},
\end{equation}
where $N_{\rm post}$ is the number of post-burn-in steps per walker.
The normalised Monte Carlo standard error of the posterior mean was then
\begin{equation}
    \frac{\mcse}{\sigma}
    =
    \frac{1}{\sqrt{N_{\rm eff}}}.
\end{equation}
We report the largest $\mcse/\sigma$ and the smallest $N_{\rm post}/\tau$ across parameters and amplitudes in Table~\ref{tab:posterior-convergence}.
The nonlinear posterior \map\ was obtained independently by numerical optimisation.

\subsection{Convergence results}

\begin{table*}[t]
    \centering
    \caption{Numerical convergence tests for the posterior calculations.
    The local DALI posterior is evaluated within the selected Fisher radius
    $R=46$.
    The selected grid $N=2301$ is compared with the refined grid
    $N=4601$.
    The MAP tolerance is determined by the finite grid spacing.
    The MCMC diagnostics quantify the sampling precision of the nonlinear
    posterior mean.}
    \label{tab:posterior-convergence}
    \begin{tabular}{lrrrrr}
        \toprule
        Amplitude
        & $d_{\rm F}(\bar{\theta}_{4601}-\bar{\theta}_{2301})$
        & $d_{\rm F}(\hat{\theta}_{4601}-\hat{\theta}_{2301})$
        & $\epsilon_{\rm MAP}$
        & max.\ MCSE$/\sigma$
        & min.\ $N_{\rm post}/\tau$ \\
        \midrule
        $A=0$ & $1.29\times10^{-10}$ & $0.00\times10^{0}$ & $3.16\times10^{-2}$ & $6.35\times10^{-3}$ & $774.3$ \\
        $A=1$ & $6.71\times10^{-12}$ & $0.00\times10^{0}$ & $3.16\times10^{-2}$ & $6.21\times10^{-3}$ & $809.1$ \\
        $A=2$ & $1.47\times10^{-11}$ & $0.00\times10^{0}$ & $3.16\times10^{-2}$ & $6.16\times10^{-3}$ & $823.8$ \\
        $A=3$ & $7.35\times10^{-11}$ & $0.00\times10^{0}$ & $3.16\times10^{-2}$ & $6.14\times10^{-3}$ & $829.2$ \\
        $A=4$ & $2.49\times10^{-10}$ & $0.00\times10^{0}$ & $3.16\times10^{-2}$ & $6.17\times10^{-3}$ & $822.0$ \\
        $A=5$ & $8.46\times10^{-10}$ & $0.00\times10^{0}$ & $3.16\times10^{-2}$ & $6.17\times10^{-3}$ & $820.3$ \\
        $A=6$ & $1.21\times10^{-9}$ & $0.00\times10^{0}$ & $3.16\times10^{-2}$ & $5.92\times10^{-3}$ & $890.3$ \\
        $A=7$ & $3.81\times10^{-9}$ & $0.00\times10^{0}$ & $3.16\times10^{-2}$ & $6.15\times10^{-3}$ & $825.0$ \\
        $A=8$ & $1.10\times10^{-8}$ & $0.00\times10^{0}$ & $3.16\times10^{-2}$ & $6.00\times10^{-3}$ & $867.0$ \\
        $A=9$ & $1.71\times10^{-8}$ & $0.00\times10^{0}$ & $3.16\times10^{-2}$ & $5.91\times10^{-3}$ & $893.5$ \\
        $A=10$ & $1.69\times10^{-8}$ & $0.00\times10^{0}$ & $3.16\times10^{-2}$ & $6.00\times10^{-3}$ & $867.8$ \\
        $A=11$ & $5.94\times10^{-8}$ & $0.00\times10^{0}$ & $3.16\times10^{-2}$ & $5.97\times10^{-3}$ & $877.1$ \\
        $A=12$ & $1.41\times10^{-7}$ & $0.00\times10^{0}$ & $3.16\times10^{-2}$ & $6.05\times10^{-3}$ & $852.9$ \\
        $A=13$ & $3.30\times10^{-8}$ & $2.00\times10^{-2}$ & $3.16\times10^{-2}$ & $5.93\times10^{-3}$ & $889.1$ \\
        $A=14$ & $5.36\times10^{-7}$ & $2.00\times10^{-2}$ & $3.16\times10^{-2}$ & $5.86\times10^{-3}$ & $909.1$ \\
        $A=15$ & $7.26\times10^{-8}$ & $2.00\times10^{-2}$ & $3.16\times10^{-2}$ & $5.92\times10^{-3}$ & $891.8$ \\
        $A=16$ & $1.05\times10^{-6}$ & $2.00\times10^{-2}$ & $3.16\times10^{-2}$ & $5.81\times10^{-3}$ & $927.0$ \\
        $A=17$ & $8.14\times10^{-7}$ & $0.00\times10^{0}$ & $3.16\times10^{-2}$ & $5.78\times10^{-3}$ & $934.9$ \\
        $A=18$ & $3.19\times10^{-6}$ & $0.00\times10^{0}$ & $3.16\times10^{-2}$ & $5.76\times10^{-3}$ & $940.5$ \\
        $A=19$ & $1.77\times10^{-6}$ & $2.00\times10^{-2}$ & $3.16\times10^{-2}$ & $5.79\times10^{-3}$ & $933.3$ \\
        $A=20$ & $4.52\times10^{-6}$ & $2.00\times10^{-2}$ & $3.16\times10^{-2}$ & $5.62\times10^{-3}$ & $988.1$ \\
        \midrule
        \multicolumn{6}{c}{Worst-case convergence} \\
        \addlinespace
        DALI mean & $4.52\times10^{-6}$ & --- & --- & --- & $A=20$ \\
        DALI MAP & --- & $2.00\times10^{-2}$ & $3.16\times10^{-2}$ & --- & $A=15$ \\
        MCMC mean & --- & --- & --- & $6.35\times10^{-3}$ & $A=0$, $u$ \\
        MCMC chain & --- & --- & --- & --- & $774.3$ ($A=0$, $u$) \\
        \bottomrule
    \end{tabular}
\end{table*}

The convergence results are summarised in Table~\ref{tab:posterior-convergence}.
Between the selected and refined \dali\ grids, the largest posterior-mean change was $\max d_\fish(\meantheta_{4601}-\meantheta_{2301})=4.516\times10^{-6}$, well below the adopted Fisher-distance tolerance of $0.01$.
The largest \map\ convergence change was $\max d_\fish(\maptheta_{4601}-\maptheta_{2301})=2.000\times10^{-2}$, compared with the finite-grid tolerance $\epsilon_{\map}=3.162\times10^{-2}$.
Both posterior quantities therefore satisfied the adopted numerical convergence criteria over the tested amplitude range.

For the nonlinear posterior, the worst \mcmc\ diagnostics were $\max(\mcse/\sigma)=6.353\times10^{-3}$ and $\min(N_{\rm post}/\tau)=7.743\times10^{2}$, both occurring at $A=0$ for parameter $u$.
The local \dali\ grid calculations satisfied the adopted convergence criteria, and the estimated \mcmc\ sampling precision supported the posterior comparisons presented here.

The validation scripts, numerical settings, and generated convergence outputs are provided as supplemental material with the preprint.

\section{Posterior shift validation}
\label{app:posterior-validation}

We validate the analytic expressions for the \dali\ \map\ and posterior mean systematic shifts against direct numerical calculations of the local \dali\ posterior.
The comparison is performed for the nonlinear benchmark over the full amplitude range $A=0,\ldots,20$, considering the $\alpha$ and $\beta$ systematic directions individually and their common amplitude combination.
The numerical posterior is evaluated using the integration settings validated in Appendix~\ref{app:numerical-validation}.

For each case, we compare the analytic prediction with the corresponding numerical \dali\ result.
For the posterior mean, we additionally evaluate the systematic expansion numerically, providing a direct test of the analytic reduction separately from the underlying perturbative expansion.
Parameter-wise accuracy is reported using absolute fractional errors in the individual shift components, while the discrepancy of the full shift vector is quantified by the Fisher distance defined in Eq.~\eqref{eq:fisher_distance}.
The fractional errors can become artificially large when an individual numerical shift component approaches or crosses zero, so the Fisher distance provides the more robust measure of the overall discrepancy in these cases.

The \map\ comparison is summarised in Table~\ref{tab:posterior-map-comparison}.
Across the full amplitude range, the analytic \dali\ \map\ differs from the numerical \dali\ reference by at most $0.0908$ in Fisher distance.
By contrast, the Fisher discrepancy grows to $23.1$ for the combined $\alpha+\beta$ systematic at $A=20$, corresponding to $\ddata=5$.
At this point, Fisher predicts a shift of $\Delta v=0.0260$, opposite in sign to the numerical result $\Delta v=-0.151$, whereas the analytic \dali\ prediction gives $\Delta v=-0.150$.

\begin{table*}[t]
            \centering
            \caption{Comparison of posterior \text{\scshape map} displacements for the numerical \text{\scshape dali} reference, Fisher prediction, analytic \text{\scshape dali} prediction, and \text{\scshape mcmc} posterior.}
            \label{tab:posterior-map-comparison}
        \begin{tabular}{llllrrr}
            \toprule
            \multicolumn{7}{c}{Posterior \text{\scshape map} displacement $\Delta\hat{\theta}$} \\
            \midrule
            \multicolumn{1}{c}{\begin{tabular}[c]{@{}c@{}}Systematic\end{tabular}}
            & \multicolumn{1}{c}{\begin{tabular}[c]{@{}c@{}}Systematic\\amplitude\\$A$\end{tabular}}
            & \multicolumn{1}{c}{\begin{tabular}[c]{@{}c@{}}Data\\distance\\$d_{\rm data}$\end{tabular}}
            & \multicolumn{1}{c}{\begin{tabular}[c]{@{}c@{}}Inference\\method\end{tabular}}
            & \multicolumn{1}{c}{\begin{tabular}[c]{@{}c@{}}$\Delta u$\\$(\epsilon_{\Delta u}\,[\%])$\end{tabular}}
        & \multicolumn{1}{c}{\begin{tabular}[c]{@{}c@{}}$\Delta v$\\$(\epsilon_{\Delta v}\,[\%])$\end{tabular}}
            & \multicolumn{1}{c}{\begin{tabular}[c]{@{}c@{}}Fisher\\distance\\$d_{\mathrm F}$\end{tabular}} \\
            \midrule
            $\alpha$ & $0$ & $0.000$ & numerical & $0.00000$ & $0.00000$ & $0$ \\
         &  &  & Fisher & $0.00000\;(\text{--}\%)$ & $0.00000\;(\text{--}\%)$ & $0$ \\
         &  &  & analytic & $0.00000\;(\text{--}\%)$ & $0.00000\;(\text{--}\%)$ & $0$ \\
         &  &  & \text{\scshape mcmc} & $0.00000\;(\text{--}\%)$ & $0.00000\;(\text{--}\%)$ & $0$ \\
        \addlinespace
         & $5$ & $1.123$ & numerical & $0.07185$ & $-0.01053$ & $0$ \\
         &  &  & Fisher & $0.07159\;(0.37\%)$ & $-0.00150\;(85.76\%)$ & $1.18\times10^{0}$ \\
         &  &  & analytic & $0.07185\;(0.00\%)$ & $-0.01053\;(0.02\%)$ & $2.50\times10^{-4}$ \\
         &  &  & \text{\scshape mcmc} & $0.07185\;(0.00\%)$ & $-0.01053\;(0.00\%)$ & $1.85\times10^{-6}$ \\
        \addlinespace
         & $10$ & $2.246$ & numerical & $0.14424$ & $-0.03844$ & $0$ \\
         &  &  & Fisher & $0.14318\;(0.74\%)$ & $-0.00300\;(92.20\%)$ & $4.63\times10^{0}$ \\
         &  &  & analytic & $0.14425\;(0.00\%)$ & $-0.03841\;(0.07\%)$ & $3.74\times10^{-3}$ \\
         &  &  & \text{\scshape mcmc} & $0.14424\;(0.00\%)$ & $-0.03844\;(0.00\%)$ & $1.12\times10^{-5}$ \\
        \addlinespace
         & $15$ & $3.369$ & numerical & $0.21718$ & $-0.08274$ & $0$ \\
         &  &  & Fisher & $0.21477\;(1.11\%)$ & $-0.00450\;(94.56\%)$ & $1.02\times10^{1}$ \\
         &  &  & analytic & $0.21721\;(0.01\%)$ & $-0.08260\;(0.16\%)$ & $1.77\times10^{-2}$ \\
         &  &  & \text{\scshape mcmc} & $0.21718\;(0.00\%)$ & $-0.08274\;(0.00\%)$ & $7.77\times10^{-6}$ \\
        \addlinespace
         & $20$ & $4.492$ & numerical & $0.29068$ & $-0.14244$ & $0$ \\
         &  &  & Fisher & $0.28636\;(1.49\%)$ & $-0.00600\;(95.79\%)$ & $1.78\times10^{1}$ \\
         &  &  & analytic & $0.29076\;(0.03\%)$ & $-0.14204\;(0.28\%)$ & $5.22\times10^{-2}$ \\
         &  &  & \text{\scshape mcmc} & $0.29068\;(0.00\%)$ & $-0.14244\;(0.00\%)$ & $7.94\times10^{-7}$ \\
        \addlinespace
        \midrule
        $\beta$ & $0$ & $0.000$ & numerical & $0.00000$ & $0.00000$ & $0$ \\
         &  &  & Fisher & $0.00000\;(\text{--}\%)$ & $0.00000\;(\text{--}\%)$ & $0$ \\
         &  &  & analytic & $0.00000\;(\text{--}\%)$ & $0.00000\;(\text{--}\%)$ & $0$ \\
         &  &  & \text{\scshape mcmc} & $0.00000\;(\text{--}\%)$ & $0.00000\;(\text{--}\%)$ & $0$ \\
        \addlinespace
         & $5$ & $1.012$ & numerical & $0.00986$ & $0.00781$ & $0$ \\
         &  &  & Fisher & $0.00986\;(0.03\%)$ & $0.00800\;(2.43\%)$ & $2.48\times10^{-2}$ \\
         &  &  & analytic & $0.00986\;(0.00\%)$ & $0.00781\;(0.00\%)$ & $2.15\times10^{-7}$ \\
         &  &  & \text{\scshape mcmc} & $0.00986\;(0.00\%)$ & $0.00781\;(0.00\%)$ & $6.74\times10^{-8}$ \\
        \addlinespace
         & $10$ & $2.025$ & numerical & $0.01973$ & $0.01524$ & $0$ \\
         &  &  & Fisher & $0.01971\;(0.06\%)$ & $0.01600\;(4.98\%)$ & $9.91\times10^{-2}$ \\
         &  &  & analytic & $0.01973\;(0.00\%)$ & $0.01524\;(0.00\%)$ & $3.33\times10^{-6}$ \\
         &  &  & \text{\scshape mcmc} & $0.01973\;(0.00\%)$ & $0.01524\;(0.00\%)$ & $9.82\times10^{-7}$ \\
        \addlinespace
         & $15$ & $3.037$ & numerical & $0.02960$ & $0.02230$ & $0$ \\
         &  &  & Fisher & $0.02957\;(0.10\%)$ & $0.02400\;(7.63\%)$ & $2.22\times10^{-1}$ \\
         &  &  & analytic & $0.02960\;(0.00\%)$ & $0.02230\;(0.00\%)$ & $1.28\times10^{-5}$ \\
         &  &  & \text{\scshape mcmc} & $0.02960\;(0.00\%)$ & $0.02230\;(0.00\%)$ & $1.01\times10^{-6}$ \\
        \addlinespace
         & $20$ & $4.049$ & numerical & $0.03948$ & $0.02898$ & $0$ \\
         &  &  & Fisher & $0.03943\;(0.13\%)$ & $0.03200\;(10.41\%)$ & $3.94\times10^{-1}$ \\
         &  &  & analytic & $0.03948\;(0.00\%)$ & $0.02898\;(0.00\%)$ & $3.80\times10^{-5}$ \\
         &  &  & \text{\scshape mcmc} & $0.03948\;(0.00\%)$ & $0.02898\;(0.00\%)$ & $9.32\times10^{-7}$ \\
        \addlinespace
        \midrule
        $\alpha+\beta$ & $0$ & $0.000$ & numerical & $0.00000$ & $0.00000$ & $0$ \\
         &  &  & Fisher & $0.00000\;(\text{--}\%)$ & $0.00000\;(\text{--}\%)$ & $0$ \\
         &  &  & analytic & $0.00000\;(\text{--}\%)$ & $0.00000\;(\text{--}\%)$ & $0$ \\
         &  &  & \text{\scshape mcmc} & $0.00000\;(\text{--}\%)$ & $0.00000\;(\text{--}\%)$ & $0$ \\
        \addlinespace
         & $5$ & $1.250$ & numerical & $0.08177$ & $-0.00529$ & $0$ \\
         &  &  & Fisher & $0.08145\;(0.40\%)$ & $0.00650\;(222.97\%)$ & $1.54\times10^{0}$ \\
         &  &  & analytic & $0.08177\;(0.00\%)$ & $-0.00528\;(0.06\%)$ & $4.46\times10^{-4}$ \\
         &  &  & \text{\scshape mcmc} & $0.08177\;(0.00\%)$ & $-0.00529\;(0.00\%)$ & $5.53\times10^{-6}$ \\
        \addlinespace
         & $10$ & $2.500$ & numerical & $0.16422$ & $-0.03312$ & $0$ \\
         &  &  & Fisher & $0.16289\;(0.81\%)$ & $0.01300\;(139.25\%)$ & $6.03\times10^{0}$ \\
         &  &  & analytic & $0.16422\;(0.01\%)$ & $-0.03307\;(0.15\%)$ & $6.58\times10^{-3}$ \\
         &  &  & \text{\scshape mcmc} & $0.16422\;(0.00\%)$ & $-0.03312\;(0.00\%)$ & $1.28\times10^{-5}$ \\
        \addlinespace
         & $15$ & $3.750$ & numerical & $0.24735$ & $-0.08203$ & $0$ \\
         &  &  & Fisher & $0.24434\;(1.22\%)$ & $0.01950\;(123.77\%)$ & $1.33\times10^{1}$ \\
         &  &  & analytic & $0.24739\;(0.02\%)$ & $-0.08179\;(0.29\%)$ & $3.09\times10^{-2}$ \\
         &  &  & \text{\scshape mcmc} & $0.24735\;(0.00\%)$ & $-0.08203\;(0.00\%)$ & $1.48\times10^{-5}$ \\
        \addlinespace
         & $20$ & $5.000$ & numerical & $0.33117$ & $-0.15055$ & $0$ \\
         &  &  & Fisher & $0.32579\;(1.63\%)$ & $0.02600\;(117.27\%)$ & $2.31\times10^{1}$ \\
         &  &  & analytic & $0.33131\;(0.04\%)$ & $-0.14985\;(0.46\%)$ & $9.08\times10^{-2}$ \\
         &  &  & \text{\scshape mcmc} & $0.33117\;(0.00\%)$ & $-0.15055\;(0.00\%)$ & $3.11\times10^{-6}$ \\
            \bottomrule
        \end{tabular}
    \end{table*}

The posterior mean comparison is summarised in Table~\ref{tab:posterior-mean-comparison}, where the closed analytic prediction and response expansion are given by Eqs.~\eqref{eq:mean_bias_main_scheme} and \eqref{eq:semi-analytic}, respectively.
The response expansion differs from the directly integrated \dali\ mean shift by at most $0.035$ in Fisher distance, while the closed analytic mean prediction reaches $1.32$ over the full amplitude range.
For the combined $\alpha+\beta$ systematic at $A=20$, corresponding to $\ddata=5$, the response expansion has a Fisher distance of $0.035$ and componentwise fractional errors of $0.02\%$ and $0.18\%$, whereas the closed analytic mean has a Fisher distance of $0.992$ and componentwise fractional errors of $2.59\%$ and $5.29\%$.
Thus, the response expansion accurately reproduces the numerical mean shifts across the tested range, using the complete unbiased \dali\ posterior as the reference distribution, while the closed analytic approximation remains a good approximation but gradually loses accuracy as the systematic mismatch increases.

\begin{table*}[t]
            \centering
            \caption{Comparison of posterior mean systematic shifts for the numerical \text{\scshape dali} reference, analytic \text{\scshape dali} prediction, systematic response expansion, and \text{\scshape mcmc} posterior.}
            \label{tab:posterior-mean-comparison}
        \begin{tabular}{llllrrr}
            \toprule
            \multicolumn{7}{c}{Posterior mean systematic shift $\Delta\bar{\theta}$} \\
            \midrule
            \multicolumn{1}{c}{\begin{tabular}[c]{@{}c@{}}Systematic\end{tabular}}
            & \multicolumn{1}{c}{\begin{tabular}[c]{@{}c@{}}Systematic\\amplitude\\$A$\end{tabular}}
            & \multicolumn{1}{c}{\begin{tabular}[c]{@{}c@{}}Data\\distance\\$d_{\rm data}$\end{tabular}}
            & \multicolumn{1}{c}{\begin{tabular}[c]{@{}c@{}}Inference\\method\end{tabular}}
            & \multicolumn{1}{c}{\begin{tabular}[c]{@{}c@{}}$\Delta u$\\$(\epsilon_{\Delta u}\,[\%])$\end{tabular}}
        & \multicolumn{1}{c}{\begin{tabular}[c]{@{}c@{}}$\Delta v$\\$(\epsilon_{\Delta v}\,[\%])$\end{tabular}}
            & \multicolumn{1}{c}{\begin{tabular}[c]{@{}c@{}}Fisher\\distance\\$d_{\mathrm F}$\end{tabular}} \\
            \midrule
            $\alpha$ & $0$ & $0.000$ & numerical & $0.00000$ & $0.00000$ & $0$ \\
         &  &  & analytic & $0.00000\;(\text{--}\%)$ & $0.00000\;(\text{--}\%)$ & $0$ \\
         &  &  & response & $0.00000\;(\text{--}\%)$ & $0.00000\;(\text{--}\%)$ & $0$ \\
         &  &  & \text{\scshape mcmc} & $0.00000\;(\text{--}\%)$ & $0.00000\;(\text{--}\%)$ & $0$ \\
        \addlinespace
         & $5$ & $1.123$ & numerical & $0.07188$ & $-0.00989$ & $0$ \\
         &  &  & analytic & $0.07146\;(0.58\%)$ & $-0.01430\;(44.63\%)$ & $5.75\times10^{-1}$ \\
         &  &  & response & $0.07188\;(0.00\%)$ & $-0.00989\;(0.00\%)$ & $1.38\times10^{-5}$ \\
         &  &  & \text{\scshape mcmc} & $0.07192\;(0.05\%)$ & $-0.00950\;(3.90\%)$ & $5.02\times10^{-2}$ \\
        \addlinespace
         & $10$ & $2.246$ & numerical & $0.14429$ & $-0.03719$ & $0$ \\
         &  &  & analytic & $0.14306\;(0.85\%)$ & $-0.04525\;(21.65\%)$ & $1.05\times10^{0}$ \\
         &  &  & response & $0.14429\;(0.00\%)$ & $-0.03720\;(0.01\%)$ & $4.75\times10^{-4}$ \\
         &  &  & \text{\scshape mcmc} & $0.14475\;(0.32\%)$ & $-0.03718\;(0.03\%)$ & $6.40\times10^{-3}$ \\
        \addlinespace
         & $15$ & $3.369$ & numerical & $0.21724$ & $-0.08090$ & $0$ \\
         &  &  & analytic & $0.21414\;(1.42\%)$ & $-0.09098\;(12.46\%)$ & $1.30\times10^{0}$ \\
         &  &  & response & $0.21724\;(0.00\%)$ & $-0.08093\;(0.04\%)$ & $3.85\times10^{-3}$ \\
         &  &  & \text{\scshape mcmc} & $0.21671\;(0.24\%)$ & $-0.08023\;(0.82\%)$ & $8.95\times10^{-2}$ \\
        \addlinespace
         & $20$ & $4.492$ & numerical & $0.29072$ & $-0.14001$ & $0$ \\
         &  &  & analytic & $0.28446\;(2.15\%)$ & $-0.14957\;(6.83\%)$ & $1.22\times10^{0}$ \\
         &  &  & response & $0.29074\;(0.01\%)$ & $-0.14014\;(0.09\%)$ & $1.75\times10^{-2}$ \\
         &  &  & \text{\scshape mcmc} & $0.29180\;(0.37\%)$ & $-0.14055\;(0.39\%)$ & $7.72\times10^{-2}$ \\
        \addlinespace
        \midrule
        $\beta$ & $0$ & $0.000$ & numerical & $0.00000$ & $0.00000$ & $0$ \\
         &  &  & analytic & $0.00000\;(\text{--}\%)$ & $0.00000\;(\text{--}\%)$ & $0$ \\
         &  &  & response & $0.00000\;(\text{--}\%)$ & $0.00000\;(\text{--}\%)$ & $0$ \\
         &  &  & \text{\scshape mcmc} & $0.00000\;(\text{--}\%)$ & $0.00000\;(\text{--}\%)$ & $0$ \\
        \addlinespace
         & $5$ & $1.012$ & numerical & $0.00987$ & $0.00791$ & $0$ \\
         &  &  & analytic & $0.00979\;(0.77\%)$ & $0.00724\;(8.47\%)$ & $8.72\times10^{-2}$ \\
         &  &  & response & $0.00987\;(0.00\%)$ & $0.00791\;(0.00\%)$ & $1.82\times10^{-8}$ \\
         &  &  & \text{\scshape mcmc} & $0.01119\;(13.38\%)$ & $0.00790\;(0.05\%)$ & $1.95\times10^{-2}$ \\
        \addlinespace
         & $10$ & $2.025$ & numerical & $0.01974$ & $0.01543$ & $0$ \\
         &  &  & analytic & $0.01959\;(0.77\%)$ & $0.01410\;(8.60\%)$ & $1.73\times10^{-1}$ \\
         &  &  & response & $0.01974\;(0.00\%)$ & $0.01543\;(0.00\%)$ & $7.06\times10^{-9}$ \\
         &  &  & \text{\scshape mcmc} & $0.02020\;(2.32\%)$ & $0.01544\;(0.02\%)$ & $6.63\times10^{-3}$ \\
        \addlinespace
         & $15$ & $3.037$ & numerical & $0.02962$ & $0.02258$ & $0$ \\
         &  &  & analytic & $0.02939\;(0.77\%)$ & $0.02061\;(8.73\%)$ & $2.57\times10^{-1}$ \\
         &  &  & response & $0.02962\;(0.00\%)$ & $0.02258\;(0.00\%)$ & $1.01\times10^{-7}$ \\
         &  &  & \text{\scshape mcmc} & $0.03069\;(3.64\%)$ & $0.02256\;(0.12\%)$ & $1.72\times10^{-2}$ \\
        \addlinespace
         & $20$ & $4.049$ & numerical & $0.03950$ & $0.02936$ & $0$ \\
         &  &  & analytic & $0.03920\;(0.77\%)$ & $0.02676\;(8.87\%)$ & $3.39\times10^{-1}$ \\
         &  &  & response & $0.03950\;(0.00\%)$ & $0.02936\;(0.00\%)$ & $3.09\times10^{-7}$ \\
         &  &  & \text{\scshape mcmc} & $0.04033\;(2.09\%)$ & $0.02931\;(0.18\%)$ & $1.56\times10^{-2}$ \\
        \addlinespace
        \midrule
        $\alpha+\beta$ & $0$ & $0.000$ & numerical & $0.00000$ & $0.00000$ & $0$ \\
         &  &  & analytic & $0.00000\;(\text{--}\%)$ & $0.00000\;(\text{--}\%)$ & $0$ \\
         &  &  & response & $0.00000\;(\text{--}\%)$ & $0.00000\;(\text{--}\%)$ & $0$ \\
         &  &  & \text{\scshape mcmc} & $0.00000\;(\text{--}\%)$ & $0.00000\;(\text{--}\%)$ & $0$ \\
        \addlinespace
         & $5$ & $1.250$ & numerical & $0.08181$ & $-0.00456$ & $0$ \\
         &  &  & analytic & $0.08129\;(0.63\%)$ & $-0.00955\;(109.60\%)$ & $6.50\times10^{-1}$ \\
         &  &  & response & $0.08181\;(0.00\%)$ & $-0.00456\;(0.00\%)$ & $2.59\times10^{-5}$ \\
         &  &  & \text{\scshape mcmc} & $0.08230\;(0.61\%)$ & $-0.00451\;(1.04\%)$ & $8.06\times10^{-3}$ \\
        \addlinespace
         & $10$ & $2.500$ & numerical & $0.16427$ & $-0.03171$ & $0$ \\
         &  &  & analytic & $0.16258\;(1.03\%)$ & $-0.04052\;(27.80\%)$ & $1.14\times10^{0}$ \\
         &  &  & response & $0.16427\;(0.00\%)$ & $-0.03171\;(0.02\%)$ & $9.16\times10^{-4}$ \\
         &  &  & \text{\scshape mcmc} & $0.16467\;(0.24\%)$ & $-0.03170\;(0.01\%)$ & $5.66\times10^{-3}$ \\
        \addlinespace
         & $15$ & $3.750$ & numerical & $0.24740$ & $-0.07994$ & $0$ \\
         &  &  & analytic & $0.24293\;(1.81\%)$ & $-0.09000\;(12.59\%)$ & $1.30\times10^{0}$ \\
         &  &  & response & $0.24741\;(0.01\%)$ & $-0.07999\;(0.07\%)$ & $7.60\times10^{-3}$ \\
         &  &  & \text{\scshape mcmc} & $0.24699\;(0.17\%)$ & $-0.07926\;(0.85\%)$ & $9.02\times10^{-2}$ \\
        \addlinespace
         & $20$ & $5.000$ & numerical & $0.33118$ & $-0.14775$ & $0$ \\
         &  &  & analytic & $0.32259\;(2.59\%)$ & $-0.15557\;(5.29\%)$ & $9.92\times10^{-1}$ \\
         &  &  & response & $0.33124\;(0.02\%)$ & $-0.14802\;(0.18\%)$ & $3.50\times10^{-2}$ \\
         &  &  & \text{\scshape mcmc} & $0.33170\;(0.16\%)$ & $-0.14781\;(0.04\%)$ & $1.28\times10^{-2}$ \\
            \bottomrule
        \end{tabular}
    \end{table*}

\section{Individual Systematic Results}
\label{app:systematic-results}

\begin{figure*}[h!]
    \centering

    \newcommand{\shiftfigwidth}{0.55\textwidth}

    \includegraphics[width=\shiftfigwidth]{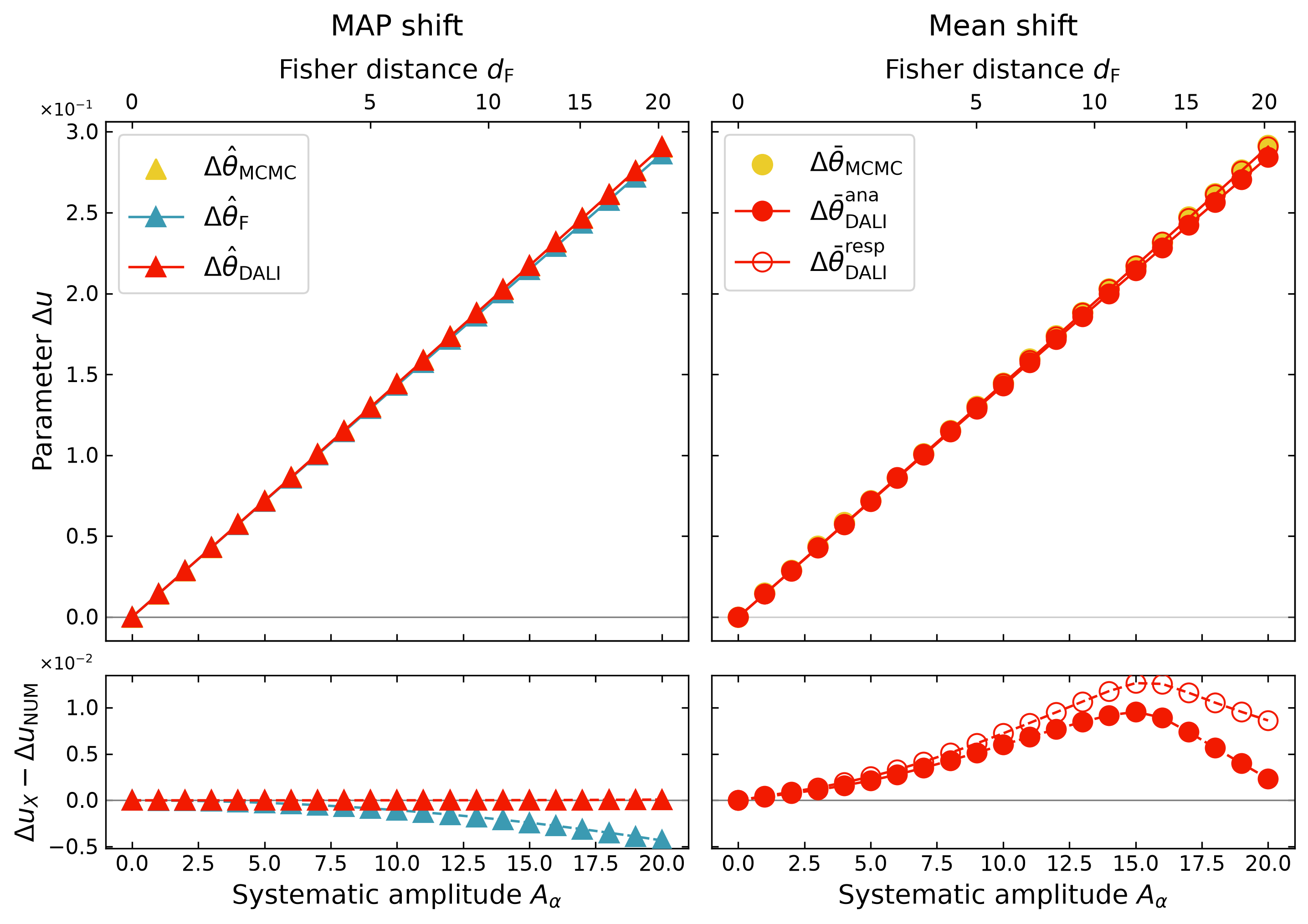}

    \small (a) Parameter $u$.

    \vspace{0.5em}

    \includegraphics[width=\shiftfigwidth]{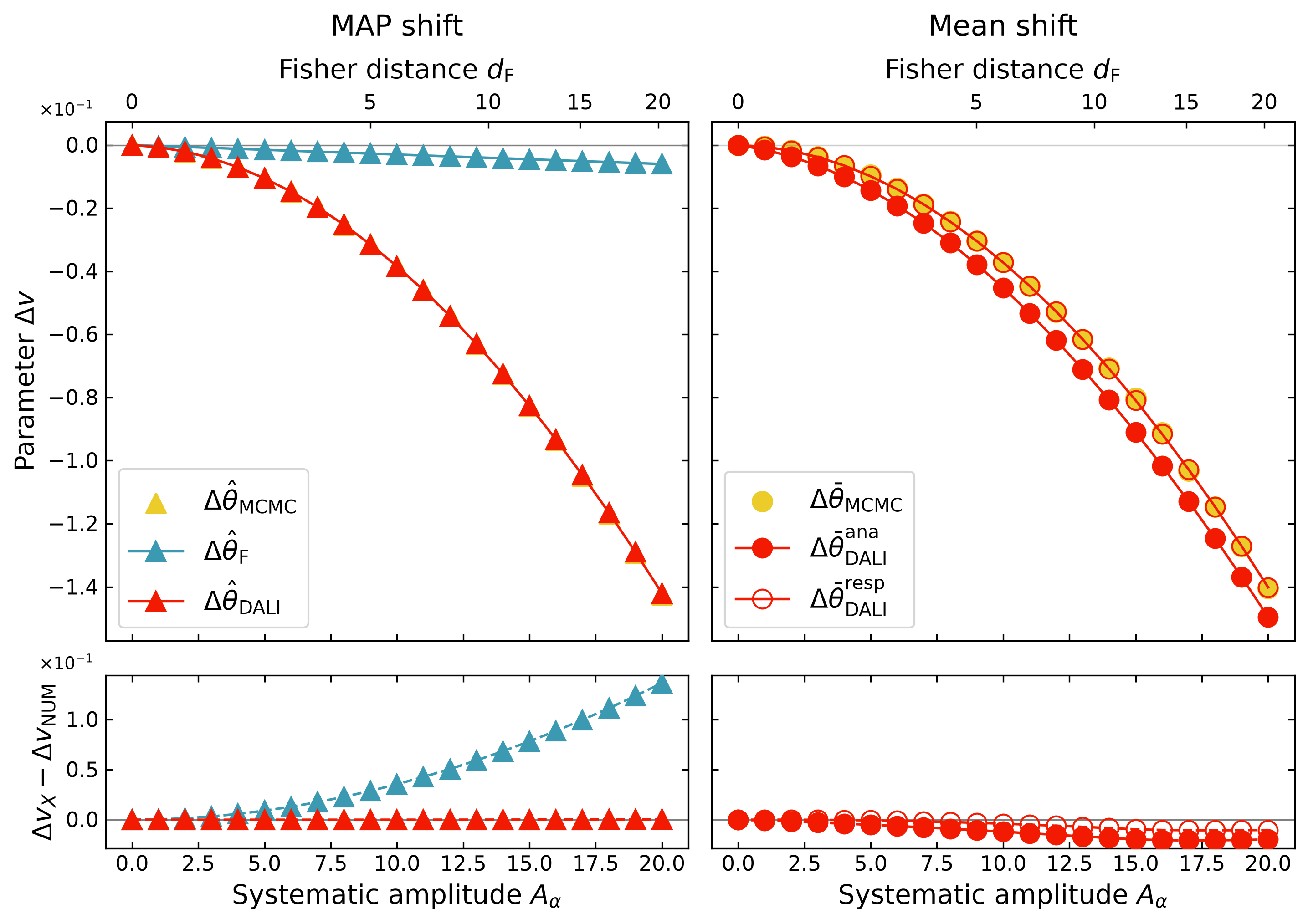}

    \small (b) Parameter $v$.

    \caption{
        Systematic \map\ and posterior mean shifts for the $\alpha$ systematic as a function of systematic amplitude.
        The upper panels compare the Fisher and \dali\ predictions with the \mcmc\ shifts, with the corresponding Fisher distance $d_{\rm F}$ shown on the upper axes.
        The lower panels show the differences between the analytic predictions and the directly evaluated numerical \dali\ \map\ and posterior mean.
    }
    \label{fig:alpha-shifts}
\end{figure*}

\begin{figure*}[h!]
    \centering

    \newcommand{\shiftfigwidth}{0.55\textwidth}

    \includegraphics[width=\shiftfigwidth]{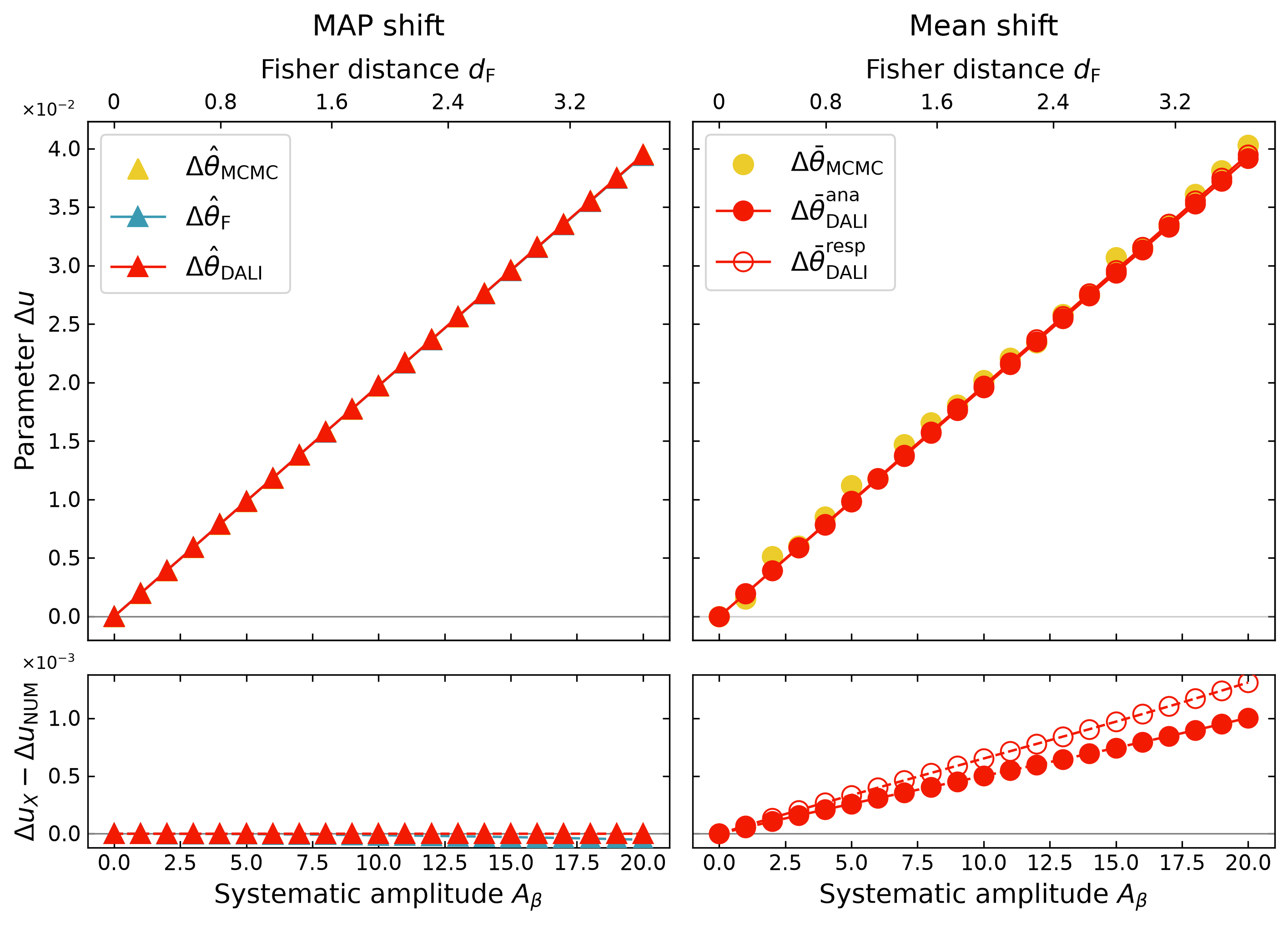}

    \small (a) Parameter $u$.

    \vspace{0.5em}

    \includegraphics[width=\shiftfigwidth]{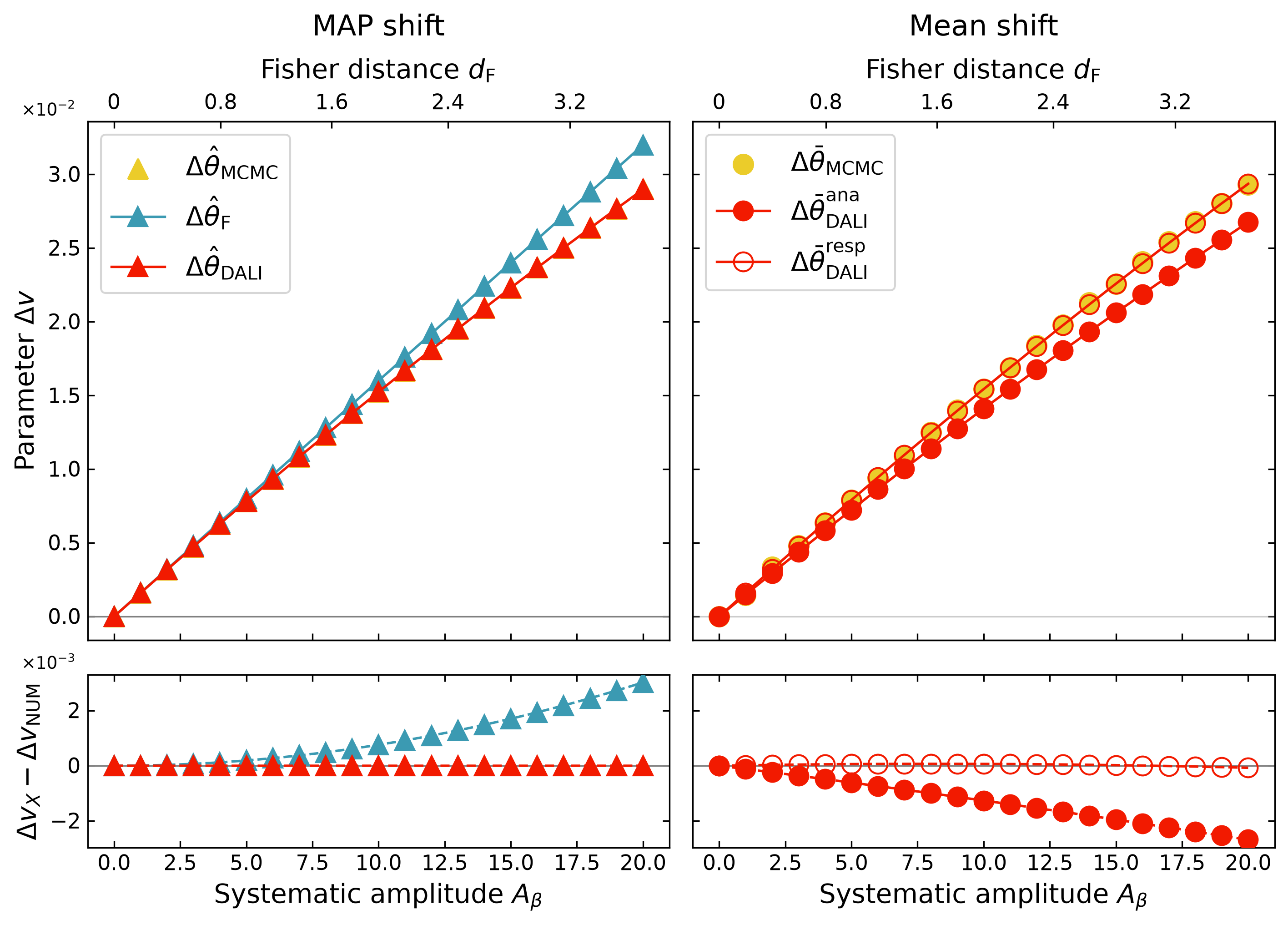}

    \small (b) Parameter $v$.

    \caption{
        Systematic \map\ and posterior mean shifts for the $\beta$ systematic as a function of systematic amplitude.
        The upper panels compare the Fisher and \dali\ predictions with the \mcmc\ shifts, with the corresponding Fisher distance $d_{\rm F}$ shown on the upper axes.
        The lower panels show the differences between the analytic predictions and the directly evaluated numerical \dali\ \map\ and posterior mean.
    }
    \label{fig:beta-shifts}
\end{figure*}

For completeness, we present the parameter-shift and posterior-reconstruction results for the individual $\alpha$ and $\beta$ systematic mismatches.
These complement the combined $\alpha+\beta$ results discussed in Section~\ref{sec:posterior_reconstruction} and allow the response to each systematic direction to be examined separately.
Figures~\ref{fig:alpha-shifts} and \ref{fig:beta-shifts} show the corresponding \map\ and posterior mean shifts as functions of systematic amplitude.

\begin{figure}[h!]
    \centering
    \includegraphics[width=0.55\textwidth]{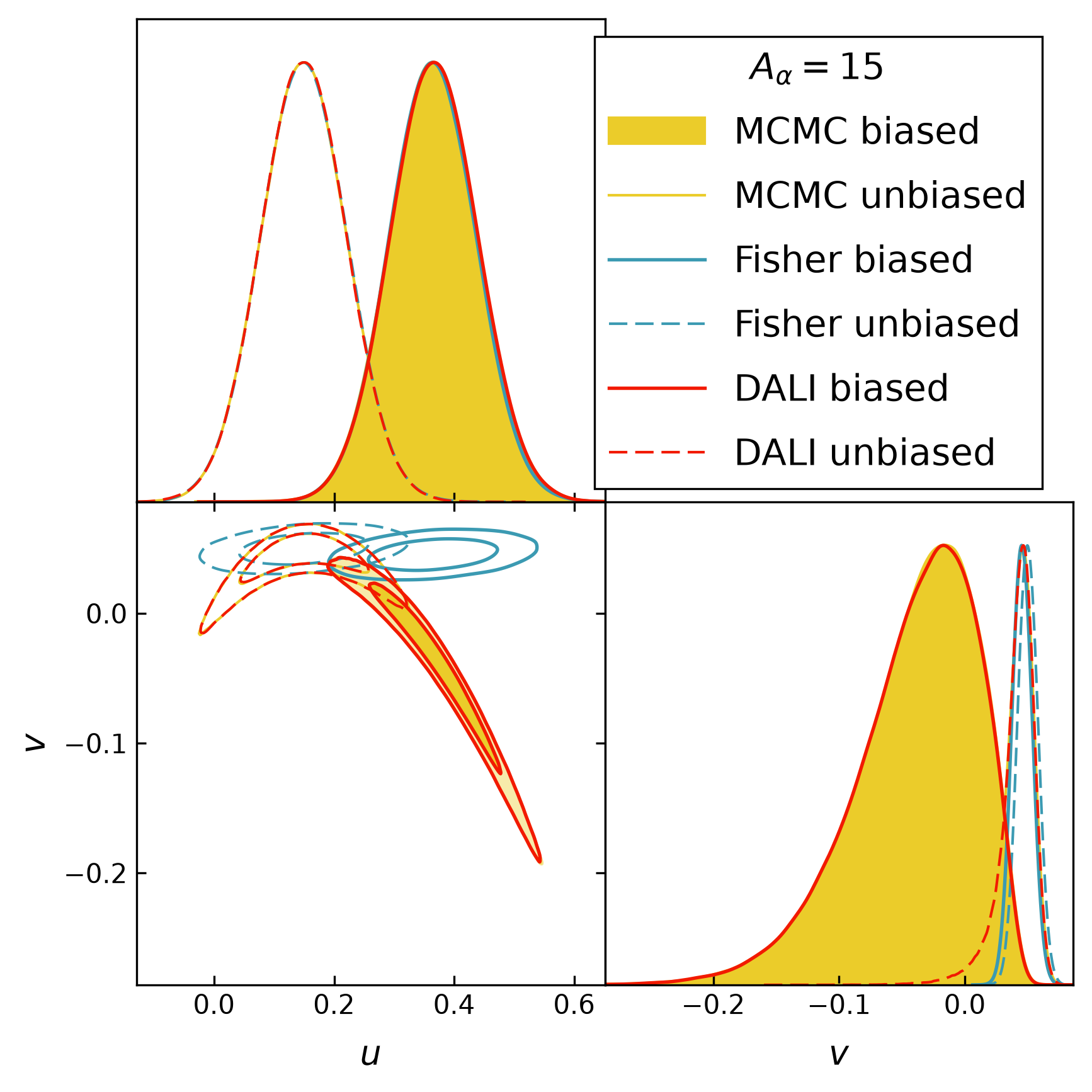}
    \caption{
        Posterior reconstruction under the individual $\alpha$ systematic mismatch at $A_\alpha=15$.
        The 1D marginalized posteriors and 2D joint posterior are shown for the \mcmc\ posterior (yellow), Fisher approximation (blue), and \dali\ reconstruction (red).
        For \mcmc, the biased posterior is filled and the unbiased reference is unfilled with solid contours; for Fisher and \dali, the biased posteriors are solid and the unbiased references are dashed, all unfilled.
        The $\alpha$ systematic produces a pronounced nonlinear deformation of the posterior as it is displaced from the unbiased reference.
        The \dali\ reconstruction captures this deformation and the resulting shift of the \mcmc\ posterior, whereas the Fisher approximation cannot reproduce the changing posterior shape.
    }
    \label{fig:alpha-posterior}
\end{figure}

\begin{figure}[h!]
    \centering
    \includegraphics[width=0.55\textwidth]{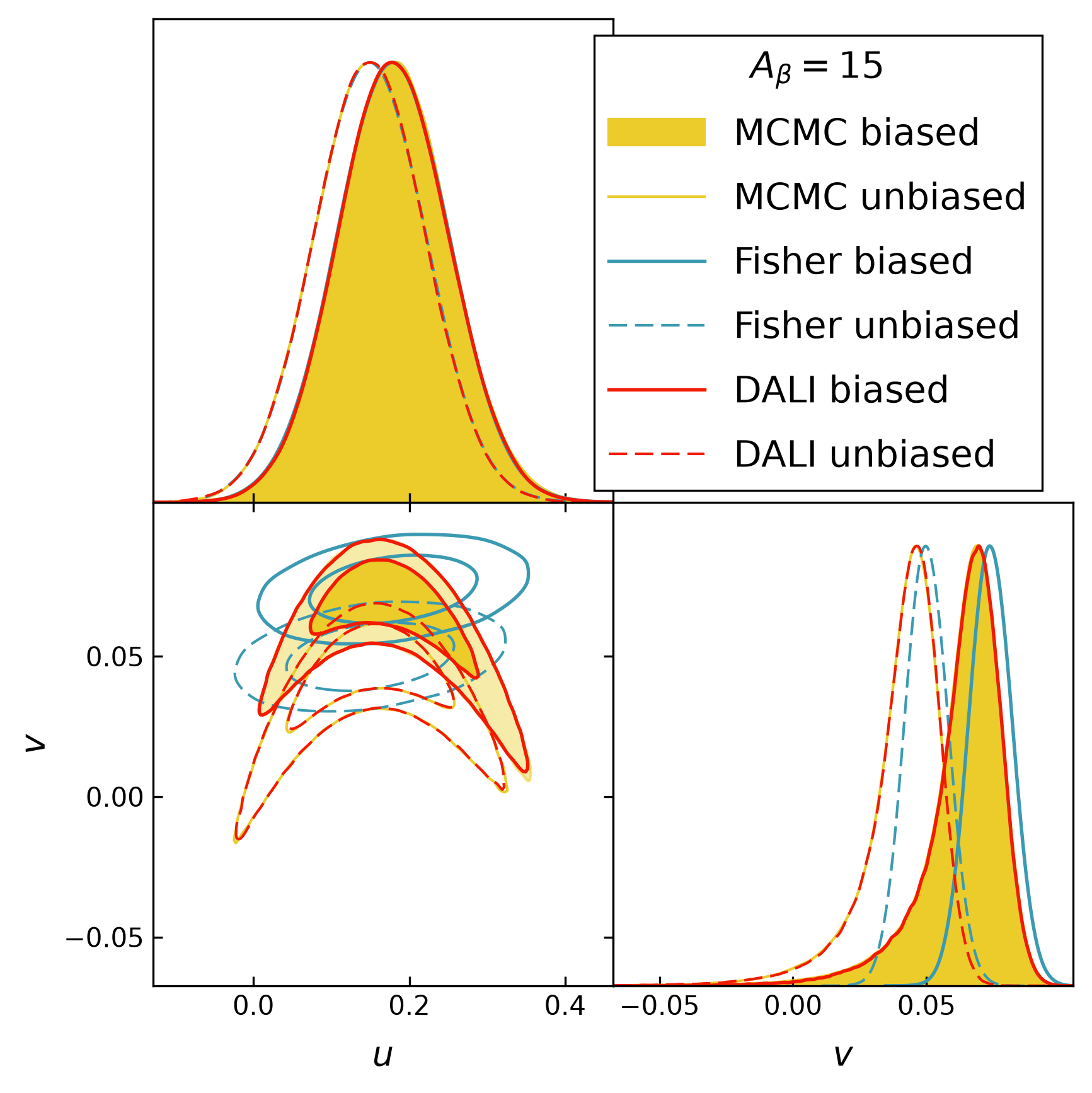}
    \caption{
        Posterior reconstruction under the individual $\beta$ systematic mismatch at $A_\beta=15$.
        The 1D marginalized posteriors and 2D joint posterior are shown for the \mcmc-sampled nonlinear posterio (yellow), Fisher approximation (blue), and \dali\ reconstruction (red).
        For \mcmc, the biased posterior is filled and the unbiased reference is unfilled with solid contours; for Fisher and \dali, the biased posteriors are solid and the unbiased references are dashed, all unfilled.
        The $\beta$ systematic primarily displaces the posterior toward larger $v$, with substantially less deformation than for the $\alpha$ direction.
        In this case, the response is therefore closer to a translation of the posterior than to a strong change in its shape.
    }
    \label{fig:beta-posterior}
\end{figure}

Figures~\ref{fig:alpha-posterior} and \ref{fig:beta-posterior} show the corresponding posterior reconstructions at $A_\alpha=15$ and $A_\beta=15$, respectively.
The $\alpha$ systematic produces a pronounced nonlinear deformation of the posterior, whereas the $\beta$ systematic primarily displaces the posterior toward larger $v$ with comparatively little change in shape.

\end{document}